\RequirePackage{fix-cm}
\documentclass[twocolumn]{svjour3_arxiv}          
\smartqed  
\usepackage{graphicx,color}
\usepackage{amssymb, amsmath,hyperref}
 \usepackage{mathptmx}     
\usepackage{latexsym}
\usepackage{enumitem}
\usepackage[switch]{lineno}

\begin{document}
\title{Built to Last: Functional and structural mechanisms in the moth olfactory 
network mitigate effects of neural injury}

\author{Charles B. Delahunt \and Pedro D. Maia \and J. Nathan Kutz}

\institute{Charles B. Delahunt  \at
\textit{(a)} Dept of Applied Mathematics, \textit{(b)} Computational Neuroscience Center; University of Washington, Seattle, WA, USA. \\
\email{delahunt@uw.edu}
\and
Pedro D. Maia \at
Department of mathematics, University of Texas at Arlington, Texas, USA. \\ 
\email{pedro.maia@uta.edu}           
\and
J. Nathan Kutz  \at
Dept of Applied Mathematics, University of Washington, Seattle, WA, USA.
\email{kutz@uw.edu}
}


\maketitle
\section*{Abstract}
Most organisms suffer neuronal damage throughout their lives, which can impair performance of core behaviors. 
Their neural circuits need to maintain function despite injury, which in particular requires preserving key system outputs.
In this work, we explore whether and how certain structural and functional neuronal network motifs act as injury mitigation mechanisms. 
Specifically, we examine how \textit{(i)} Hebbian learning, \textit{(ii)} high levels of noise, and \textit{(iii)} parallel inhibitory and excitatory connections contribute to the robustness of the olfactory system in the \textit{Manduca sexta} moth.  
We simulate injuries on a detailed computational model of the moth olfactory network calibrated to \textit{in vivo} data.
The injuries are modeled on focal axonal swellings, a ubiquitous form of axonal pathology observed in traumatic brain injuries and other brain disorders.  
Axonal swellings effectively compromise spike train propagation along the axon, reducing the effective neural firing rate delivered to downstream neurons.
All three of the network motifs examined significantly mitigate the effects of injury on readout neurons, either by reducing injury's impact on readout neuron responses or by restoring these responses to pre-injury levels.  
These motifs may thus be partially explained by their value as adaptive mechanisms to minimize the functional effects of neural injury. 
More generally, robustness to injury is a vital design principle to consider when analyzing  neural systems.  


%
%
%
%
\section*{Author Summary}
Neuronal injuries and degeneration are commonplace across species and organisms, compromising cognitive function and 
neurosensory integration.  Despite abrupt or gradual neuron impairment, neuronal circuits and networks must maintain 
functionality of key outputs in order to provide robust performance. A common type of impairment found at the cellular level in 
traumatic brain injuries and in a number of leading neurological diseases  is known as Focal Axonal Swelling (FAS). Similarly to 
demyelination,  FAS typically distorts, confuses or blocks the information encoded in spike trains. 
We simulate FAS-inspired injuries on a detailed computational model of the olfactory circuitry of the \textit{Manduca sexta} 
moth  to examine the injury mitigation effects of various common neural motifs such as high noise levels and Hebbian plasticity. 
Our results indicate that these motifs may serve as adaptive mechanisms for mitigating the effects of neuronal injury. For example, 
the Hebbian learning mechanism strongly mitigates effects of injury on system function by protecting vital downstream neurons 
from effects of upstream injury.

%
%

\section{Introduction} 

Injuries are inevitable for most organisms, yet maintaining a satisfactory level of functionality can be decisive for their survival. 
The progressive wear of a honeybee’s wings, for example, challenges the insect to sustain its load lift or face less nourishing foraging trips \cite{higginson,roberts2015}. 
Functional robustness is desirable for neural systems as well. 
While computer devices operate in a regime of near-zero tolerance for physical damage, the middle-aged human brain undergoes significant neuronal losses on a daily basis  \cite{pouget2014}. 
Robustness to injury is often overlooked when analyzing the purpose and function of neural structures while the transmission of maximum information, high signal-to-noise ratio, and low energy consumption are often primarily considered \cite{ganguli2012}.  
Analyzing neural information processing in the context of these principles is certainly important, but arguably incomplete. 
The goal of this work is to examine whether certain neural mechanisms and architectural structures can be understood as adaptive, built-in systems for robustness to brain injury from trauma, aging, and/or other disorders. 
That is, we examine how biological neural systems are ``built to last''. 
In particular, we explore how certain neural architectures can  protect the system's key downstream outputs (the ``deliverables'' of the system) from the effects of damage to upstream regions.  

\begin{figure*}[t]
\begin{center}
\includegraphics[width= .9 \textwidth]{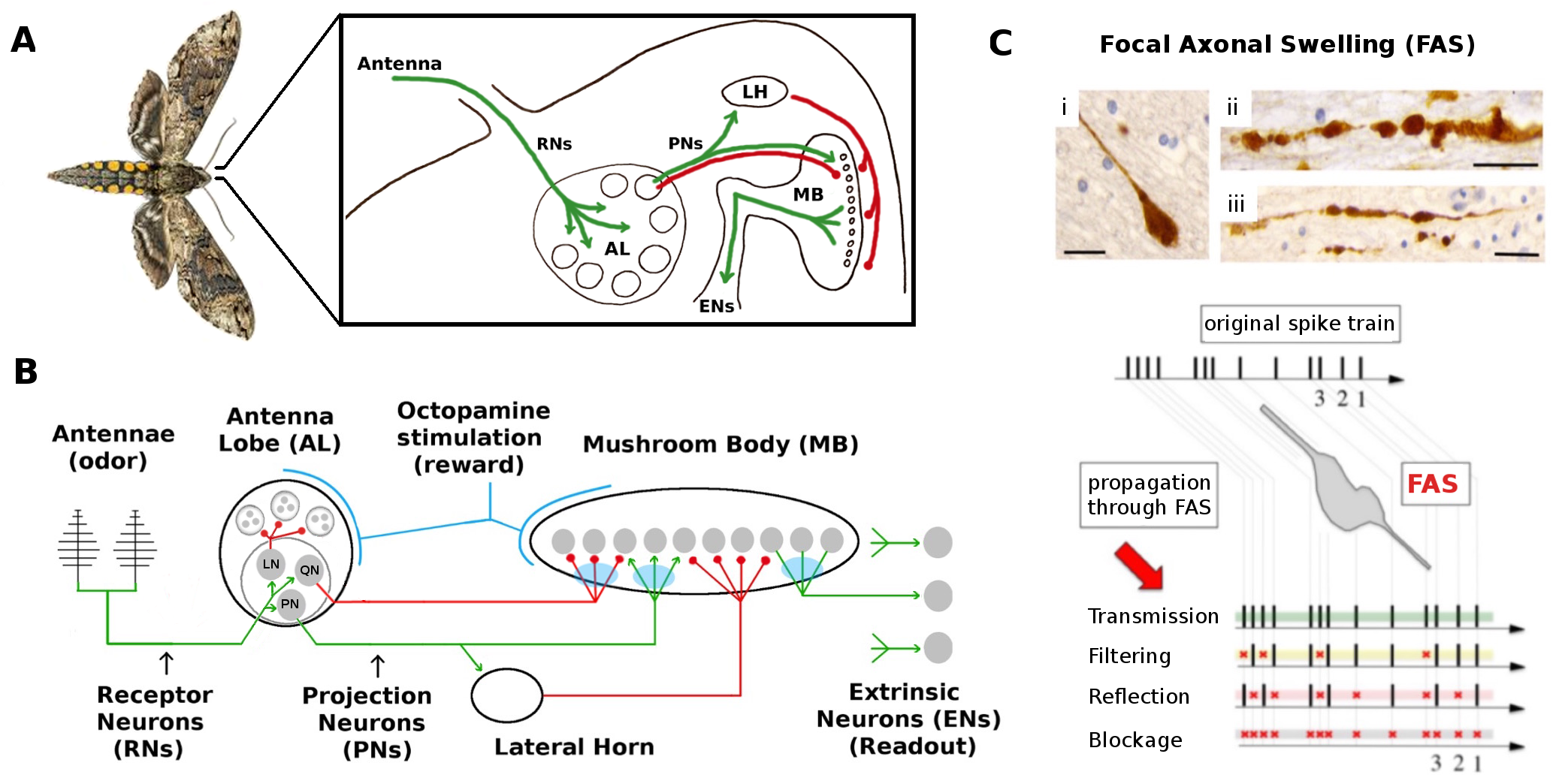}  
\caption{\small {  {\bf Overview of the Moth Olfactory Network}  
(MON) {\bf and axonal injury mechanisms. A, B:} The MON is organized as a feedforward cascade of four distinct subnetworks and a reward mechanism \cite{martin2011,kvello2009}. 
Receptor Neurons (RNs) in the Antennae detect relevant odors in the environment and transmit specific signals to the Antennal Lobe (AL) \cite{wilson2008,masse2009}, which acts as a pre-amp, providing gain control and sharpening odor codes \cite{bhandawat2007}.
The AL neurons project odor codes forward to the Mushroom Body (MB) \cite{campbellMushroomBody} by means of noisy   \cite{galizia2014} excitatory Projection Neurons (PNs), and to a smaller number of parallel inhibitory neurons (here called QNs). 
The Kenyon Cells in the MB fire sparsely (due to global inhibition from the Lateral Horn),  and encode odor signatures as memories \cite{perisse2013,honeggerTurner2011}.
Finally, Extrinsic Neurons (ENs) are viewed as readout units that interpret the MB codes, delivering actionable output to the rest of the body \cite{campbell2013,hige2015}. 
In response to reward (sugar at the proboscis), a large neuron releases octopamine in the AL and MB. 
In the AL, this neuromodulator induces stronger responses in AL neurons (stimulation), though intra-AL connections are not plastic.
Synaptic connections into and out of the MB (AL$\rightarrow$MB and MB$\rightarrow$ENs) are plastic \cite{cassenaer,masse2009} given octopamine, though octopamine does not stimulate KCs in the MothNet model (biological data is lacking).  
Learning fails in the absence of octopamine \cite{hammer1995,hammerMenzel1998}.
For accurate anatomy, see \textit{e.g.} \cite{lofaldi2010}. 
{\bf{C:}} Focal Axonal Swellings (FAS) are ubiquitous across all severities of traumatic brain injuries and present in other leading brain disorders. 
They can cause some or all neural spikes in the train to die off in transit, reducing the overall firing rate arriving at the downstream target neuron. 
Adding FAS-like effects to the MON are the basis of our damage/injury protocols. 
Panel C is adapted from \cite{maiaReactionTime}.
See the Materials and Methods section for details.   }  }
\label{introFig}
\end{center}
\end{figure*}

The olfactory system of the {\em Manduca sexta} moth, though simple, shares many neural structures and mechanisms with higher organisms  \cite{Eisthen2002,Klambt2009}.  
These include \textit{(i)} Hebbian plasticity,  \textit(ii) reward-triggered stimulation of neural outputs via neuromodulators, \textit{(iii)} high noise levels, and  \textit{(iv)} inhibitory feed-forward channels running parallel to excitatory channels. 
It is thus an ideal model organism to investigate the injury mitigation effects of these elements.
\textit{MothNet} is a computational model of this olfactory network which incorporates known biophysical parameters and which was calibrated to \textit{in vivo} firing rate data recorded during learning tasks \cite{delahuntMoth1}. 
See Figure \ref{introFig} for a system schematic. 

The moth olfactory network (MON) also contains well-defined Readout Neurons (ENs, for Extrinsic Neurons), which are downstream outputs that deliver key actionable encodings to the rest of its body \cite{campbell2013,hige2015}. 
From a functional viewpoint, internal damage is unimportant as long as the key outputs (readouts) of the system are preserved.
Thus, to examine injury mitigation effects we ran  \textit{in silico} simulations of neural injury on the \textit{MothNet} model, and measured how the firing rates (FRs) of readout ENs were affected by injuries and by injury-mitigation mechanisms.
%
%
%
\begin{table}[t]
\caption{List of acronyms used throughout this article.}
\hspace*{-0.5cm}
 \centering 
 \vspace{1em}
 \scriptsize
\begin{tabular}{ l l l l }
\hline \\
AL & Antennal Lobe  & PN & Projection Neuron \\
EN & Extrinsic (Readout) Neuron &  QN & Inhibitory Projection Neuron \\
FAS & Focal Axonal Swelling & RN & Receptor Neuron   \\
FR & Firing Rate &  SNR & Signal-to-Noise Ratio \\
MB & Mushroom Body & SSNR & Signal-to-Spontaneous Noise Ratio    \\
MON & Moth Olfactory Network &&  \\
\hline\\
\end{tabular}
\end{table}
%
 
Maia et al. \cite{Maia2019} recently introduced a computational model for the cellular level effects that may distort firing rates due to major forms of neuronal injury. 
They posit phenomenological input/output rules to transform healthy neuronal spike train responses into injured ones, with filters 
that can be either discrete-time (for spike trains) or continuous-time (for firing rates) signal processors. 
These filters were derived modeling the effects of demyelination and Focal Axonal Swellings (FAS), which are present in a broad array of neurological disorders \cite{Smith1994,Maia2014_1,Maia2014_2,Maia2015}.
 
Fig.\ref{introFig}C exemplifies how a FAS-like injury distorts the propagation of spike trains along the axon, effectively blocking or filtering signals encoded to downstream neurons.  
In this work, we are agnostic concerning the exact biological underpinnings and pathological mechanisms that may affect an injured/aging moth. 
Instead, we simply posit that its neurons might be exposed to detrimental effects that can affect their signaling capabilities.  
In this sense, and as explained in Maia et al. \cite{Maia2019}, FAS-based filters provide a more nuanced way to model neuronal malfunction than purely binary ablation which treats a neuron and/or its connections as either fully functional or 100\% impaired.
Recent computational studies that consider the effects of FAS-like injury in neural networks are providing new insight to decision-making deficits \cite{maiaReactionTime}, learning impairments \cite{Lusch2018,Rudy2016}, memory deterioration \cite{Weber2017}, and motor-function decline \cite{Kunert2017}. 
 
While FAS models effects at the level of spike trains, it also has a meaningful representation in Firing Rate models such as \textit{MothNet}. 
In particular, unlike ablation, FAS causes reduced but still non-zero FRs. 
In addition, the low-pass filtering effect of FAS, which impacts closely-bunched clusters of spikes more than sparse spikes, in analogous manner impacts high FRs more strongly than low FRs. 
Thus FAS, applied in a FR model, results in neuron FRs being reduced but not ablated, with high FR neurons affected more strongly than low FR neurons.
For a fixed amount of total damage, FAS results in relatively many partially-damaged neurons, while ablation results in relatively few destroyed neurons. 

In our simulations, we  varied the parameters of each network feature-under-test,  applied FAS-type injuries to different subnetworks of the system (simulating the outcome of a traumatic brain injury or concussion) and assessed the net effects on EN outputs. 
In particular, we examined two aspects of  FR behavior for a single representative  EN: \textit{(i)} changes in raw FR, and \textit{(ii)} the ability of the EN to discriminate between a trained and untrained odor. 
 
Our experiments led to four main findings concerning injury-mitigation structures in the moth olfactory system: 
\begin{enumerate}
\item The learning mechanism, based on  the combination of octopamine stimulation and Hebbian growth, can restore both the magnitude and discriminative ability of downstream readout neurons after upstream neurons are injured.\\
\item The presence of inhibitory neurons parallel to excitatory neurons connecting the subnetworks can mitigate the effects of injury through a ``canceling out" effect. \\ 
\item A broad noise envelope on neural firing rates (FRs) enables the strongest downstream neural responses to continue to exceed action-triggering thresholds, despite upstream injury. \\
\item Simple ablation injuries in upstream region produce distinct downstream effects from those of more biologically plausible types of injury. That is, ablation may be a poor proxy for naturalistic injury in some neural systems.\\
\end{enumerate}

Concerning item 1 (learning as  an injury mitigation mechanism), we note that while it is intuitive that Hebbian plasticity might help repair an injured network,  it is not clear that Hebbian updates \textit{alone} can repair damage. 
Hebbian ``fire together, wire together'' updates are proportional to the FRs of both the incoming and the receiving neurons. 
Hebbian plasticicity thus requires that the upstream neurons have sufficiently strong FRs, but precisely these FRs are impaired by injury.
Our experiments indicate that octopamine-induced stimulation of the injured upstream neurons is crucial to post-injury plasticity, because it boosts the FRs of upstream, injured neurons back to levels that enable non-trivial Hebbian updates. 
Without such stimulation, the FRs of the injured neurons, as well as the FRs induced by the injured neurons in the downstream neurons, are too low to induce gains  in synaptic strength via a Hebbian mechanism \cite{delahuntMoth1}.
 
Our computational approach allowed us to quantify the mitigating effect of a neural structure-under-test  as a function of injury level, injury location, and structure parameters.  
We recognize that, as in any computational model, specific quantitative outcomes necessarily depend on the particular parameters and assumptions of the \textit{MothNet} model.  
However, \textit{MothNet}'s architecture is tethered to known biophysical findings, and its parameter values are calibrated to biophysical findings and \textit{in vivo} FR data \cite{delahuntMoth1}.
We believe this approach enables our experimental results to refer back meaningfully to the biological structure.  
  

\section{Results}

\begin{figure*}[ t!]
\begin{center}
\includegraphics [width=.85\textwidth] {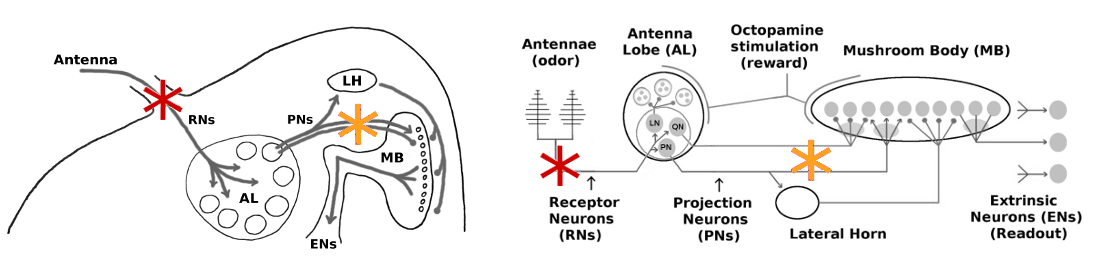}  
\caption{\small {  {\bf Location of injuries in experiments.}  Damage to Antennae affects  the Receptor Neurons (RNs)  (red stars) and reduce the overall input to the Antennal Lobe (AL). Damage to the  AL$\rightarrow$MB channel (orange stars) will weaken the signals passed by both excitatory projection neurons (PNs) and inhibitory projection neurons (QNs). 
  }  }
\label{injuryRnsPns}
\end{center}
\end{figure*}

Throughout this work, we target two distinct regions with our injury protocols: 
\textit{(i)} the Antennae and \textit{(ii)} the channel between the Antennal Lobe (AL) and the Mushroom Body (MB). 
Their specific locations are shown in Fig.\ref{injuryRnsPns}.

\textit{(i)} The antennae comprise the outermost region of the olfactory system and are arguably the most exposed to external environmental shocks. Damage in this location should affect primarily the Receptor Neuron (RN) subpopulation ($\sim$30,000). 
We note that  hundreds of RNs responsive to a given odor are spread throughout the antennae, ensuring that localized damage to an antenna does not disproportionately reduce the response to a particular odor.

\textit{(ii)} The AL$\rightarrow$MB channel is an internal region and is one of the core centers for signal transfer in the network. 
Damage in this location should affect both excitatory projection neurons (PNs) and inhibitory projection neurons (QNs) that link the AL to the MB. 

\begin{figure} [hb!]
\begin{center}
\includegraphics[width= .45\textwidth]{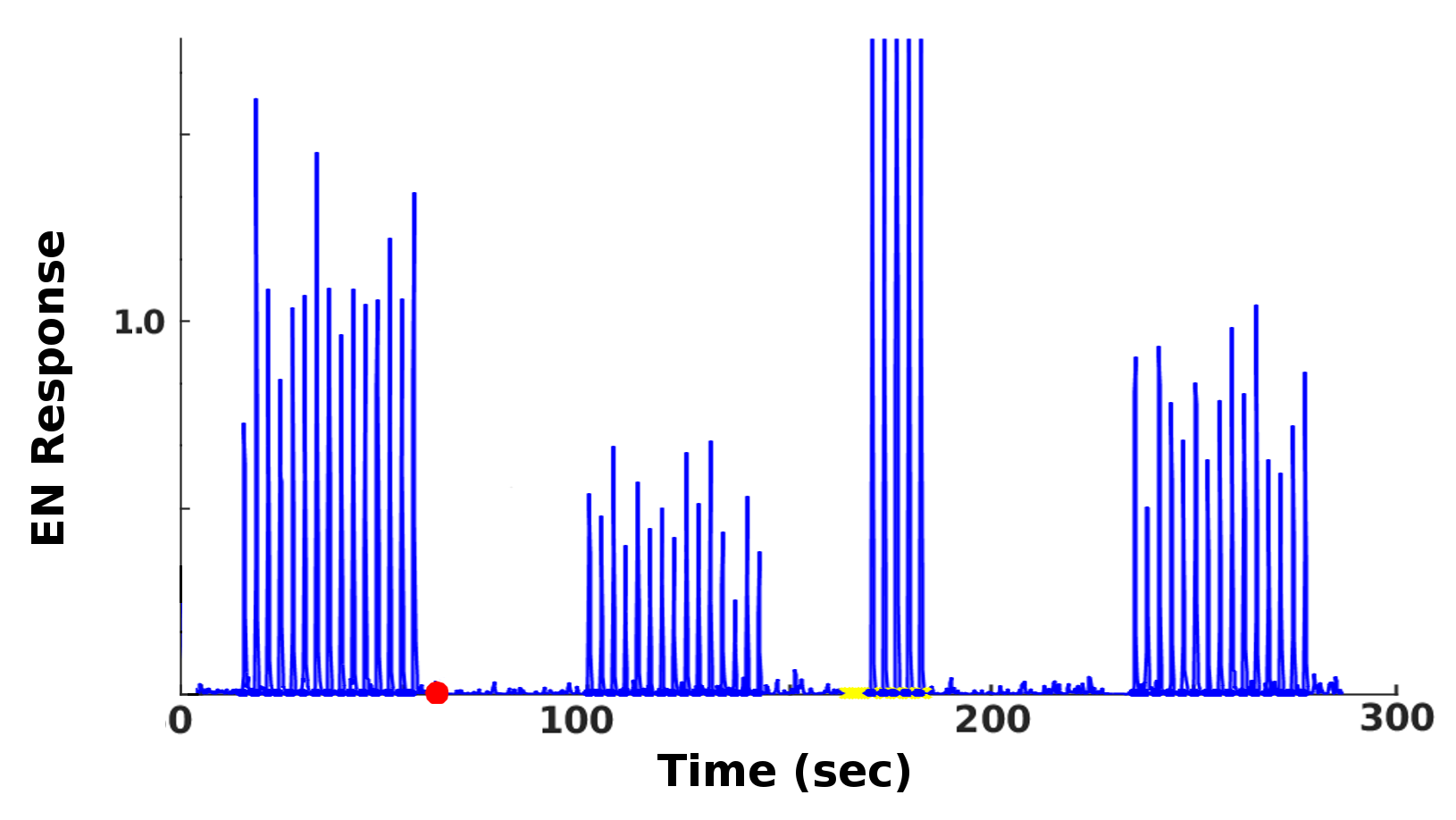}  
\caption{\small {  {\bf Typical EN timecourse.}  
Readouts from the EN in a typical experiment, in which injury attenuated the EN odor response, and training partly restored it.
\textit{x}-axis is time.
\textit{y}-axis gives magnitude of EN response (dimensionless units), with pre-injury response to the odor  $\approx$ 1 (absent odor, EN response $\approx$ 0). }
Events (with times in parentheses) are: Naive response (20-55); injury (red dot at 60); injured response (100-150); 5 puffs training (170-190), with the very strong (cropped) responses  due to octopamine; post-training response (240-280).
  }  
\label{typicalEnTimecourse}
\end{center}
\end{figure}

The Moth Olfactory Network (MON) contains some plastic synaptic connections, and it can learn~\cite{delahuntMoth1}: In response to reward (sugar at the proboscis), a large neuron sprays octopamine over the AL and MB. 
This strengthens the plastic synaptic connection in the AL$\rightarrow$MB  and MB$\rightarrow$Readout channels in a Hebbian-like way, enabling readout neurons (extrinsic neurons, ENs) to deliver actionable encodings to the rest of the body. 
Typical EN responses before injury, after injury, and after subsequent training, are shown in Fig.\ref{typicalEnTimecourse}.


%

\subsection{Plasticity-induced recovery from injury} \label{F1experiments}
  
The goal of this set of experiments was to examine how far the Hebbian learning mechanism can compensate for neural injury.
In the first experiment, RNs in the Antennae$\rightarrow$AL channel were injured (Fig.\ref{injuryRnsPns}A).
In the second experiment, PNs in the AL$\rightarrow$MB channel were targeted (Fig.\ref{injuryRnsPns}B).
AL noise was set to naturalistic levels (calibrated per \textit{in vivo} data~\cite{delahuntMoth1}) and FAS-like injury levels ranged from 0\% to 60\%. 
The MON was subsequently retrained with 5 odor puffs, close to sufficient to max out the allowable synaptic weights.
The average EN readout response was recorded, as a key measure of the actionable output of the system. 
A typical timecourse is shown in Fig.\ref{typicalEnTimecourse}.
In each experiment, over 30 (\textit{n} = 34 - 38, mean = 36) MothNet instances were generated from template (\textit{i.e.} a specification of network parameters used to randomly generate MothNet instances) and tested at each injury level.
We examined two properties of the EN readout: 

\textit{(i)} Magnitude,  a basic property relevant to triggering behavioral response to an odor. 
Examining effects on EN magnitude required no pre-training of the network prior to injury.

\textit{(ii)} Discriminative ability between two odors (one trained and one control), measured as the Fisher linear discriminant 
\begin{equation}
~~~~~~~~~~Fd =  \frac{ \mu_{train} - \mu_{control} } { 0.5( \sigma_{train} + \sigma_{control} ) }
 \end{equation} 
 where $\mu, \sigma$ are the mean and std. dev. of EN responses to trained and control odors.
In these simulations, prior to injury one odor was trained so the  system could discriminate the trained odor vs. control odor ($Fd \approx 5$).
We either used two randomly-generated odor profiles  with broad, overlapping projections onto the AL; or \{odor +  noise\} vs.  noise, with mean noise magnitude set between 0.2 and 1.0 times the odor magnitude (in  \textit{MothNet}, odor magnitude is controlled by a scalar which multiplies unit-length odor vectors  before they are inputted to the RNs).

\begin{figure}[bh]
\begin{center}
\includegraphics[width=  .45\textwidth]{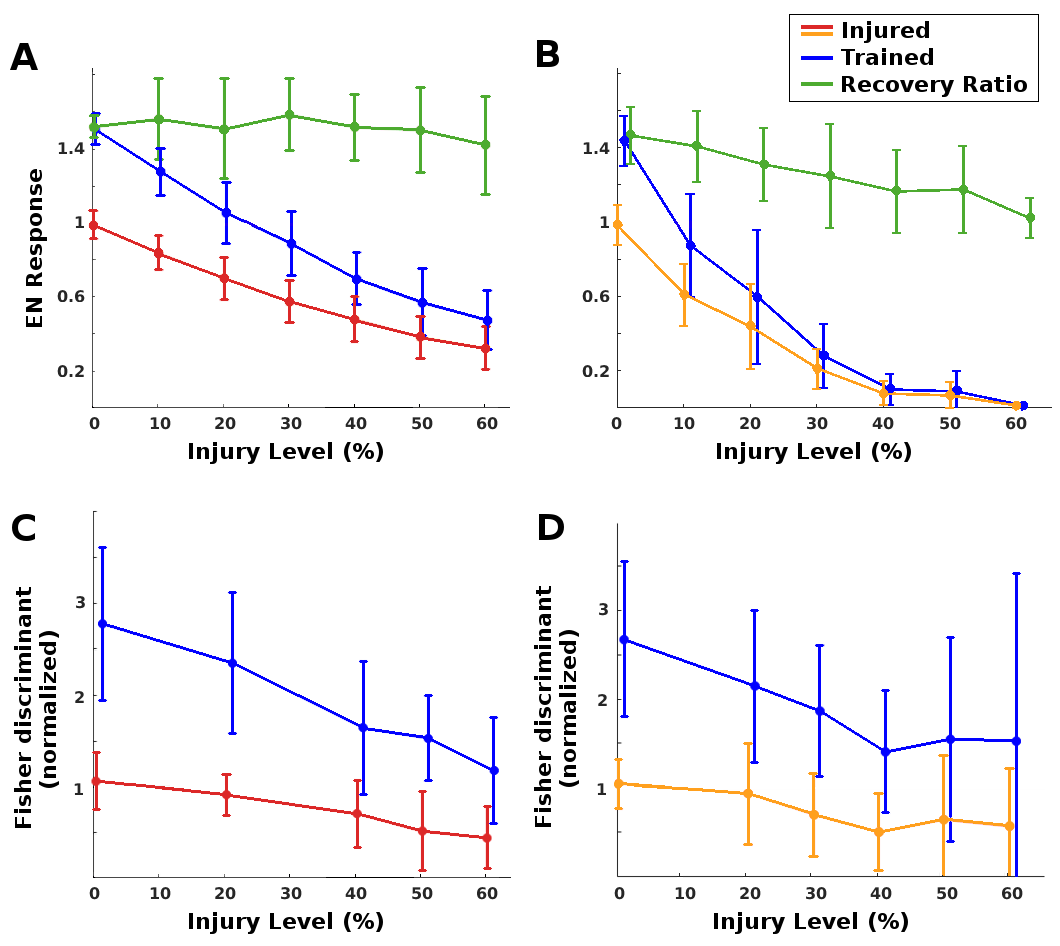}  
\caption{\small {  {\bf Learning as injury compensation mechanism.} 
Red/orange: Post-injury EN odor response, normalized by naive, healthy odor response. 
Blue: Post-training EN response, normalized by naive, healthy odor response. 
Green: Relative increase from post-injury response due to training. 
$\mu \pm \sigma$.
{\bf{A:}} Injury to RNs: Trained EN responses (blue) fully regained their pre-injury levels 
(black line) from injured levels (red) if injury was on average $\leq$ 25\%.
The ability of training to recover lost ground was fairly steady \textit{vs} injury level (green).
{\bf{B:}} Injury to PNs was more traumatic: Post-injury EN response (orange) was lower, 
and trained responses (blue) fully regained pre-injury levels if injury was on average $\leq$ 8\%.
Also, the ability of training to recover lost ground decreased as injury level increased (green).
Each datapoint shows the mean and std dev, over  $n$ = 31 to 62 (mean = 40) moth instances. 
A moth's EN response was defined as its mean response to 15 odor exposures.
{\bf{C, D:}} Changes in Fisher discriminant between pre-trained odor and control odor, due to injury and subsequent additional training (values are normalized to the initial Fisher discriminant). 
Injury reduced the ability to discriminate (red/orange curves), while post-injury training fully restored it to above baseline (blue curves).
For 0\% injury, slight deviations from 1 are due to variations in responses to two groups of odor puffs, pre- and post-injury.
C: Injury to RNs. D: Injury to PNs.
  }  }
\label{learningResultsPanel}
\end{center}
\end{figure}

\textit{(i)} EN magnitude: As expected, injury reduced raw EN response magnitude, and training restored some of this loss. 
The MON was much more robust to RN (antennae) damage than to PN damage. 
Complete restoration was achieved (on average) for injury levels below 25\% for RN damage and below 8\% for PN damage. 
At these injury values, injury reduced EN odor responses to approximately 70\% of the naive baseline, and training restored them to baseline. 
See the plots in Fig.\ref{learningResultsPanel}(A, B). 
We note that at low injury levels, the system was able to boost EN output by about 140\% to 150\%, a value constrained by the model's saturation parameter for the synaptic connection weights. 
At high levels of injury to PNs, however,  the learning mechanism's ability to recover EN performance decreased  (see green curves in Fig.\ref{learningResultsPanel}). 

\textit{(ii)} Discrimination: Injury had less effect on discriminative ability than on magnitude, since injury affected both trained odor and control EN responses. 
Given two odors, injury reduced discrimination, while retraining readily restored all losses (Fisher discriminant plots in Fig.\ref{learningResultsPanel}(C, D).
Between \{odor + noise\} vs. noise, injury had no effect on discriminative ability at any noise level, likely because the sparsely-firing MB is an effective noise filter \cite{delahuntMoth1}.
Post-injury training served to further increase discrimination between \{odor + noise\} and noise (results not shown).

\subsection{Inhibitory neurons and protective canceling out effect} \label{F2experiments}

Each glomerulus in the AL has $\approx$ 5 excitatory PNs that feed forward to the MB.
The moth also has a smaller number of inhibitory neurons (here called QNs) that also feed forward to the MB, analogous to and in parallel with the PNs.
We note that these feed-forward QNs are one of three inhibitory networks in the AL-MB.
The other two, \textit{viz.}   lateral inhibitory neurons within the AL \cite{wilson2005}  and global sparsity-inducing inhibition onto the MB from the Lateral Horn  \cite{bazhenovStopfer2010} (or global self-inhibition by the MB as in \textit{drosophila} \cite{lin2014}), have different functions and are assumed to be non-plastic.
Our experiments target the  QNs, which innervate only a subset of MB neurons, and are presumed in \textit{MothNet} to be plastic like PNs. 

The goal of this set of experiments was to test whether the existence of QNs parallel to PNs might  mitigate the effect of injuries applied to this region (orange stars in Fig.\ref{injuryRnsPns}). 
We varied the QN:PN ratio  (0, 2, 4, 5, and 7 QNs/glomerulus, compared to 5 PNs/glomerulus) while injuring the AL$\rightarrow$MB channel. 
Each parameter pair (\textit{e.g.} ``4 QNs, 50\% FAS injury'') had at least 30 moth instances (31 to 40, mean = 35).
We found out that higher numbers of QNs correlated strongly with reduced effects on EN output magnitudes from upstream injury, but had no clear effect on discriminative ability.
Results reported here are for effects on EN magnitude.
 
\begin{figure}[bh!]
\begin{center}
\includegraphics[width= .5 \textwidth]{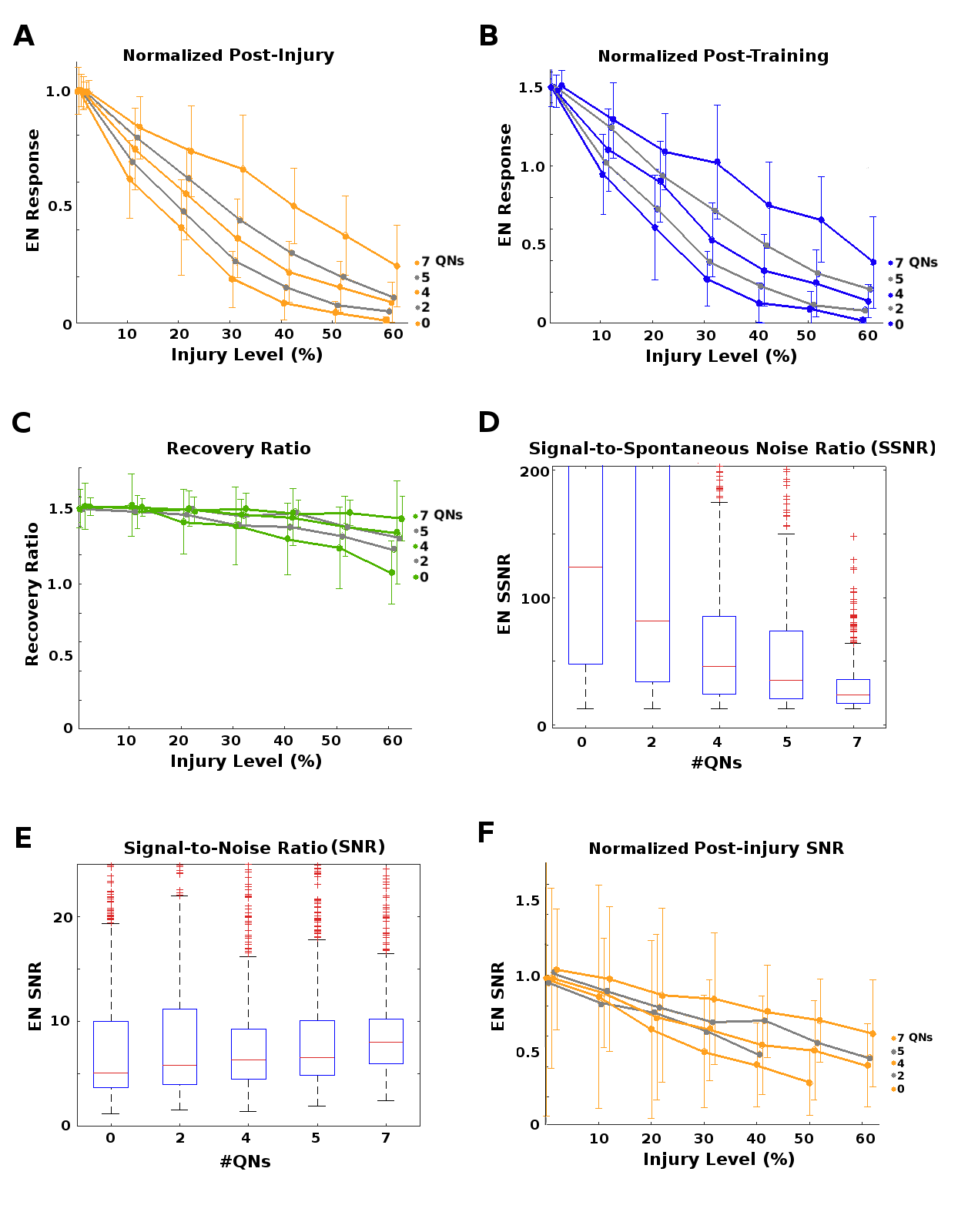}  
\caption{\small {  {\bf Effects of parallel inhibitory channels.}  
{\bf{A:}}  Post-injury EN odor responses normalized by naive healthy odor responses,  \textit{vs} injury level. 
Each curve corresponds to a number of QNs per 5 PNs, from 0 to 7.
Higher QN:PN ratios resulted in much lower impact on EN responses for a given level of injury.
{\bf{B:}}  Post-training EN odor responses   normalized by naive healthy odor responses,   \textit{vs} injury level. 
Each curve corresponds to a number of QNs per 5 PNs, from 0 to 7.
Higher QN:PN ratios resulted in stronger recovery.
{\bf{C:}}  Ratio of post-training to post-injury EN odor responses  \textit{vs} injury level. 
Recovery rate dropped off at injury levels $\geq$ 20\% for \#QN = 0, but higher numbers of QNs reduced this drop-off, \textit{i.e.} ensured better recovery. 
{\bf{D:}} Box-whisker plots (showing 25 and 75\%iles as a blue box and median as a red line) of the ratio of naive healthy EN odor responses to spontaneous EN noise (SSNR). This measure of signal clarity was much lower in moths with high QN counts.
{\bf{E:}} Raw Signal-to-Noise Ratio (SNR)  of naive healthy EN responses was fairly uniform across \#QNs.
{\bf{F:}} Post-injury SNR normalized by pre-injury SNR. 
High QN counts gave strong protection against injury-induced degradation of SNR.
  }  }
\label{varyQnResultsPanel}
\end{center}
\end{figure}

Moths with high QN counts had stronger post-injury EN odor responses, and post-injury training sessions allowed 
them to fully recover from much higher levels of injury than moths with few or no QNs ($\approx$ 8\% injury for QNs = 0, 
$\approx$ 15\% injury when QNs = 4, and  $\approx$ 30\% injury when QNs = 7). High QN counts had another, unexpected 
advantage regarding the Signal-to-Noise Ratio, 
\begin{equation}
~~~~~~~~~~~~~~ SNR = \mu(F) / \sigma(F),
\end{equation} 
where $\mu$ and $\sigma$ correspond to the mean and standard deviation, and  $F = \{f_i, i = 1...n\}$  is the set of discrete EN responses (peak FR) to a series of odor puffs.
Naive SNR values (\textit{i.e.} pre-injury, pre-training) were similar for all QN counts (Fig.\ref{varyQnResultsPanel}D).
While post-injury SNR always dropped proportionally to the severity of the injury,  high QN counts substantially reduced these losses to SNR, suggesting that raw EN firing rates were better preserved (Fig.\ref{varyQnResultsPanel}E). 

However, high QNs counts also carried a downside. They had a much lower EN Signal-to-Spontaneous Noise Ratio, 
\begin{equation}
~~~~~~~~~~~~~~SSNR = \mu(F)/ \mu(Sp),
\end{equation}
where $F$ is defined as above and $Sp$ is the spontaneous EN firing rate. 
The SSNR measures the clarity of the signal with respect to background noise, and their values for different QN counts 
are shown in Fig.\ref{varyQnResultsPanel}F. Many templates with high QN counts were rejected due to untenably 
high naive spontaneous noise.

\textit{P}-values corresponding to Fig.\ref{varyQnResultsPanel} (A, B) are given in Tables \ref{tableQNsPostInjury} and \ref{tableQNsPostTrain} in the Appendix. 
They indicate that the injury-mitigating effect of high QN ratios was meaningful  (we avoid the term ``significant" in association with \textit{P}-values, following the arguments in \cite{rePvalues}).

Our results demonstrate that the presence of parallel inhibitory neurons help protect the signal from injury (at the cost of decreased SSNR). 
We hypothesize that QNs  achieve this by a canceling out mechanism: When inhibitory QNs are injured, the overall transmitted signal increases, offsetting the decreases caused by injury to excitatory PNs.

\subsection{AL noise preserves the highest EN responses} \label{F3experiments}
The AL is a noisy network. We ask whether this neural noise has injury-mitigation benefits. 
We suppose that vital odor-related behavior is triggered when a discrete EN response $f$ exceeds some threshold, and that due to AL noise these responses $f$ to a particular odor (+ concentration) vary as if drawn from  a distribution. 
The moth gets $n$ exposures to a given odor plume, and thus has $n$ discrete responses $F = \{f_i, i = 1...n\}$.
Then to induce the behavior, a triggering response  (\textit{i.e.} $f_i > $ threshold) is  needed for only some, not all, $f_i$. 

In this case, it suffices for the system to protect only the strongest (top-scoring) EN responses from injury-induced attenuation in order  to maintain its behavioral response.
The goal of this experiment was to examine whether higher AL noise levels might preferentially protect the top-scoring EN responses from injury-induced attenuation.
Noting that $F$  parametrizes a Gaussian  

\begin{equation}
~~~~\mathcal{N}(\mu(F), \sigma(F)) = \mathcal{N}(mean(F), ~std~dev(F))
\end{equation}
we define this top-scoring tranche as those responses at the top end of the distribution: $\{ f_i \in F~ | ~f_i > \mu(F) + \sigma(F) \}$.
This corresponds to $f_i$ boosted by fortuitous noise effects and thus most likely to exceed the triggering threshold.
 
\begin{figure}[bh!]
\begin{center}
\includegraphics[width= .5\textwidth]{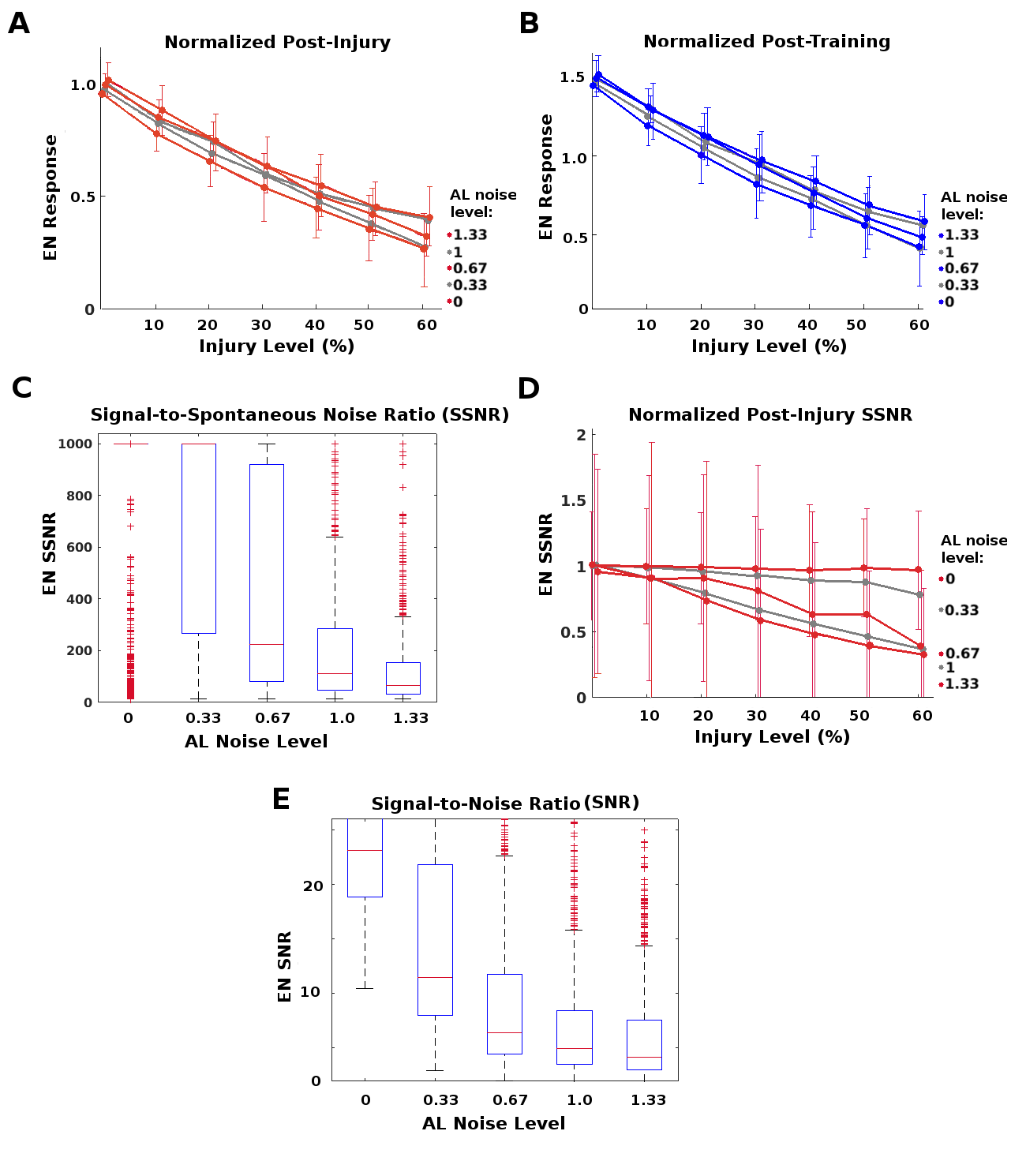}  
\caption{\small {  {\bf Effects of AL noise:} Given RN injury, AL noise protects downstream neurons from 
loss (A, B), but exacts a cost in terms of signal-to-spontaneous noise ratio (SSNR, C and D) and signal-to-noise ratio (SNR, E). 
{\bf{A:}}  Post-injury EN responses of the top 15\% tranche (\textit{i.e.} the strongest odor responses), normalized by their pre-injury response,  \textit{vs} injury level. 
Higher AL noise reduced attenuation due to injury, at any level of injury.  
Each curve corresponds to a level of AL noise, from 0 to 1.33 where 1 = ``natural" level. 
Pre-injury response = black line.
(For 0\% injury, slight deviations from 1 are due to variation in response to two groups of odor puffs, pre- and post-injury.)
{\bf{B:}}   Post-injury EN responses of the top 15\% tranche, normalized by their pre-injury response,  \textit{vs} injury level.
Higher AL noise allowed training to give full recovery of these top EN responses from larger injuries,
 $\approx$28\% injury given maximum noise \textit{vs}  $\approx$20\% injury given no AL noise.
{\bf{C:}} Naive healthy ratio of EN SSNR was much lower at high AL noise levels.
{\bf{D:}} Post-injury SSNR, normalized by pre-injury ratios, \textit{vs} injury level.  
Injury lowered SSNR far more in moths with high AL noise. 
{\bf{E:}} Naive healthy SNR  by AL noise level. SNR was much lower in moths with high AL noise. 
}  }
\label{varyALNoiseResultsPanel}
\end{center}
\end{figure}

\begin{figure}[bh!]
\includegraphics  [width=  .5\textwidth] {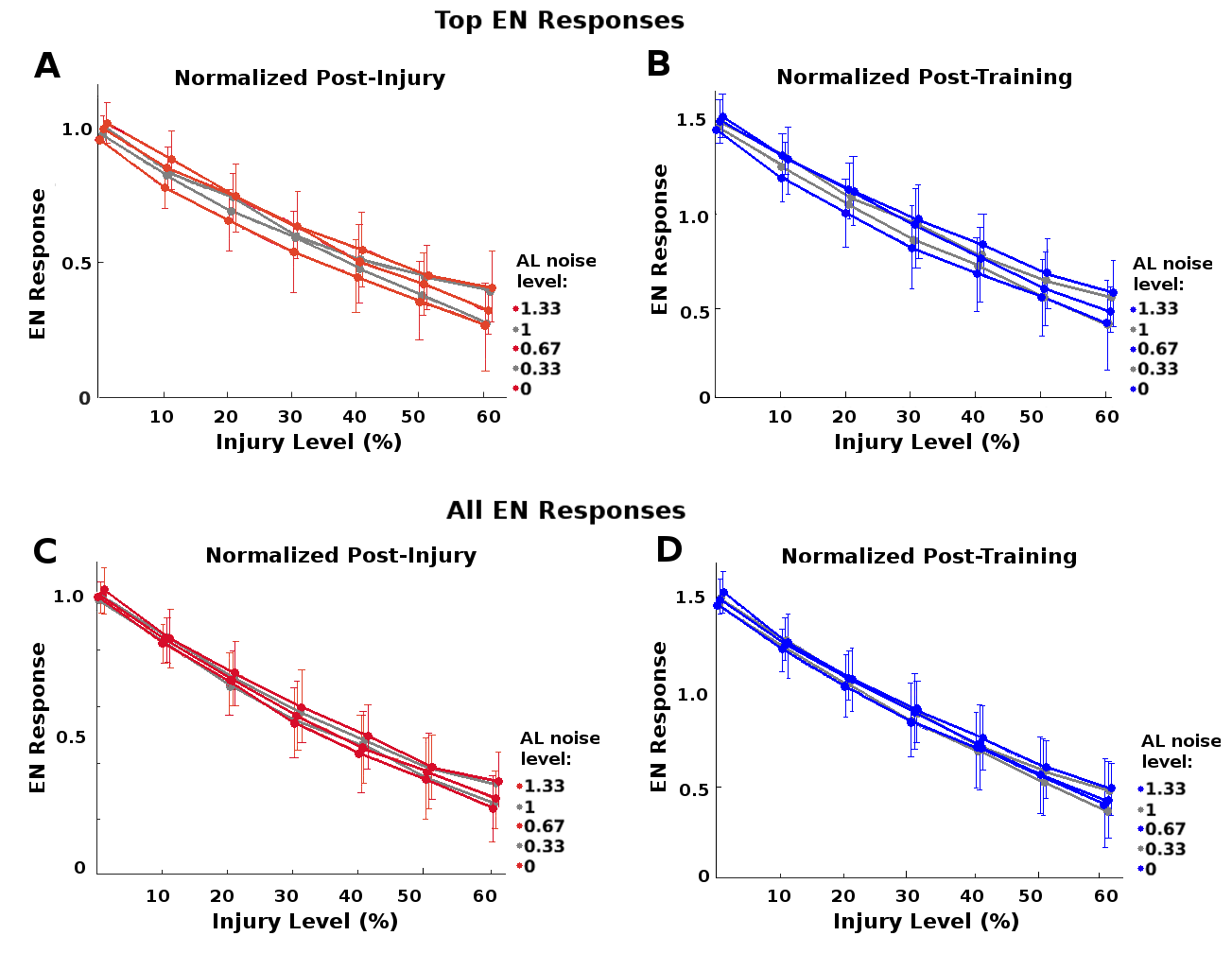} 
\caption   {\small{  {\bf Protective effects of AL noise on strongest \textit{vs} average EN responses}
Given RN injury, increased AL noise had a greater protective effect on the top 15\% tranche of EN odor responses 
than on more average odor responses, both post-injury and post-training. Each curve corresponds to a noise level. 
A wider spread of curves indicates greater injury mitigation from higher noise.
{\bf{A, B:}} Top 15\% of EN responses, normalized to their pre-injury responses, post-injury (red, grey) and post-training (blue, grey).
The spread of curves indicates the relative benefit of higher noise. 
(These are the same subplots as in Fig.\ref{varyALNoiseResultsPanel} A, B.)
{\bf{C, D:}} All responses, normalized by their pre-injury responses, post-injury (red, grey) and post-training (blue, grey).
Average EN responses had less injury mitigation benefit from high noise than top-scoring EN responses.  
 } }
\label{alNoiseMuVsMuPlusSigma}
\end{figure}

The AL noise level is controlled by a single parameter in \textit{MothNet}.
We adjusted neural noise in the AL to different multiplicative factors of the ``natural" AL noise level (\textit{i.e.} the level matching our \textit{in vivo} data).
Factors were 0, 0.33, 0.67, 1.0, and 1.33, where 1.0 is the natural level.
Various severities of FAS-like injury were applied to RNs in the Antennae$\rightarrow$AL channel (Fig.\ref{injuryRnsPns}A).
Over 30 (31 to 62, mean = 40) moth instances were generated from template for each \{AL noise, injury level\} datapoint.
To measure attenuation in top-scoring responses, we  defined the Top-End Preservation $P$ as: 

\begin{equation}
~~~~~~~~~~P(F_ j)= \frac{\mu(F_j) + \sigma(F_j)}{\mu(F_h) + \sigma(F_h)},
\end{equation}
where $F_{h}$ is the set of  pre-injury (healthy) responses to repeat applications of some stimulus, and $F_j$ is the set of discrete responses  to the same stimulus post-injury at level $j$.   
$P(F_j)$ measures how much an injury affects the top-scoring responses (represented by $\mu(F) + \sigma(F)$) when it shrinks the entire response distribution from $\mathcal{N}(\mu(F_h), \sigma(F_h))$ to $\mathcal{N}(\mu(F_j), \sigma(F_j))$.  
$P(F, j)$ ranges between 1 and 0, where 1 implies no injury-induced attenuation, and 0 implies total attenuation.
$P(F_ j))$  is plotted in Figs.\ref{varyALNoiseResultsPanel} (A, B) and \ref{alNoiseMuVsMuPlusSigma}, for both post-injury $P$ and post-injury-plus-training $P$. 

Higher AL noise increased the top-end preservation $P$ of EN responses caused by a given level of injury (Fig.\ref{varyALNoiseResultsPanel}A).
It also increased the post-training recovery possible: 
For example, full recovery occurred for injury $\leq 20\%$ when AL noise = 0, \textit{vs} $\leq 28\%$ when AL noise was greater than natural level (Fig.\ref{varyALNoiseResultsPanel}B).

However, high AL noise levels had a significant downside, namely, lower SNR (signal to noise ratio) and SSNR (signal to spontaneous noise ratio) values, as seen in Fig.\ref{varyALNoiseResultsPanel} (C-E). 
This suggests that the moth must make a trade-off between robustness to injury and signal quality. 

In addition, we found that this protection did not apply to all EN responses: 
Top-scoring EN responses received more injury-mitigation benefit from higher AL noise levels than did average EN responses, \textit{i.e.} $P(F) > \frac{\mu(F_j)}{\mu(F_h)}$.
That is, the extra robustness to injury conferred by higher noise levels was greater for top-scoring responses than for average responses.
This meshes with the notion that the system needs not protect all responses, just the ones most likely to exceed triggering threshold.
Fig.\ref{alNoiseMuVsMuPlusSigma} shows this difference in protective effect, top-scoring \textit{vs} average. 

\textit{P}-values are given in the Appendix as follows:
\textit{P}-values  corresponding to Figs.\ref{varyALNoiseResultsPanel} (A, B) are given in Tables \ref{tableALNoiseTopPostInjury} and \ref{tableALNoiseTopPostTrain}. 
\textit{P}-values corresponding to Fig.\ref{alNoiseMuVsMuPlusSigma} (C, D) are given in Tables \ref{tableALNoiseAvePostInjury} and \ref{tableALNoiseAvePostInjury}. 
\textit{P}-values comparing injury-mitigating effects on top-scoring \textit{vs} average responses, (\textit{i.e.} Fig.\ref{alNoiseMuVsMuPlusSigma}, A\,\textit{vs}\,C and B\,\textit{vs}\,D) are given in Tables \ref{tableALNoiseTopVsAllPostInjury} and \ref{tableALNoiseTopVsAllPostTrain}. 
They indicate that \textit{(i)}  the increased protective effect due to increased AL noise was meaningful on the top-scoring responses; \textit{(ii)} the protective effect was noticeably lower for average responses; and \textit{(iii)} the protective effect was meaningfully greater for top-scoring than for average responses.

Results given here are for experiments that tracked injury's effects on magnitude of EN response.
Injury's effects on odor discrimination were only slightly affected by varying AL noise, since both trained odor and control were attenuated (results not shown).


\subsection{Ablation is a poor proxy to biological injury}\label{F4experiments}
Neuronal pathologies are often modeled in a binary way, \textit{i.e.} by treating a neuron and/or its connections as either fully functional or fully impaired, and ablation injuries are widely studied in theoretical and experimental settings. 
But recent FAS studies  show  that most injured neurons maintain some residual firing rate activity. 
On large, homogeneous populations of neurons where outputs are pooled, such as the 30,000 RNs in the moth AL-MB, one can expect an approximate equivalence in ablation and FAS modulo a conversion factor.
This is because the effect of injuring or ablating any single RN is relatively small, and the overall effects of injury to the population can be approximated by average injury ratios.
In this case, we estimate that ablation alone is roughly 1.75$\times$ more harmful than FAS-like injury, \textit{i.e.} ablating $n$\%  of neurons in a population causes the same relative drop in total summed FRs as FAS injury to  $\approx1.75n$\% of the neuron population.
For calculations, see section \ref{sectionRatioCalculation}). 

However, where neuron numbers are smaller and neural outputs are not pooled, so that individual neurons have relatively unique effects on the system, it is not clear that ablation effects can be reliably mapped to effects of more biologically-plausible FAS-like injuries. 
Injury to PNs more closely resembles this case, since there are only 60 glomeruli in the moth AL, each handling unique information. 

In ablation studies, injury levels are typically measured as percentage of neurons ablated. 
To assess whether ablation is a good proxy for naturalistic FAS injury,  
we ran experiments to test whether the impacts on EN response magnitude of ablation injury vs. FAS-like injury had a consistent 1.75$\times$ relationship at these two locations, \textit{i.e.} the RN channel (red stars in Fig.\ref{injuryRnsPns}) and the PN channel (orange stars in Fig.\ref{injuryRnsPns}). 
All parameters were generated from a \textit{MothNet} template, with AL noise at natural levels and number of QNs = 0 (QNs = 2 gave similar results). 
Half the moths were injured by ablation and half were injured by FAS, with injury levels from 0 to 60\%, in order to compare the relative empirical effects on EN outputs. 
In each experiment, over 30 (31 to 40, mean = 35) moth instances were generated for each injury \{type, level, location\} datapoint. 

\begin{figure}[bh!] 
\begin{center}
\includegraphics[width=  .5\textwidth] {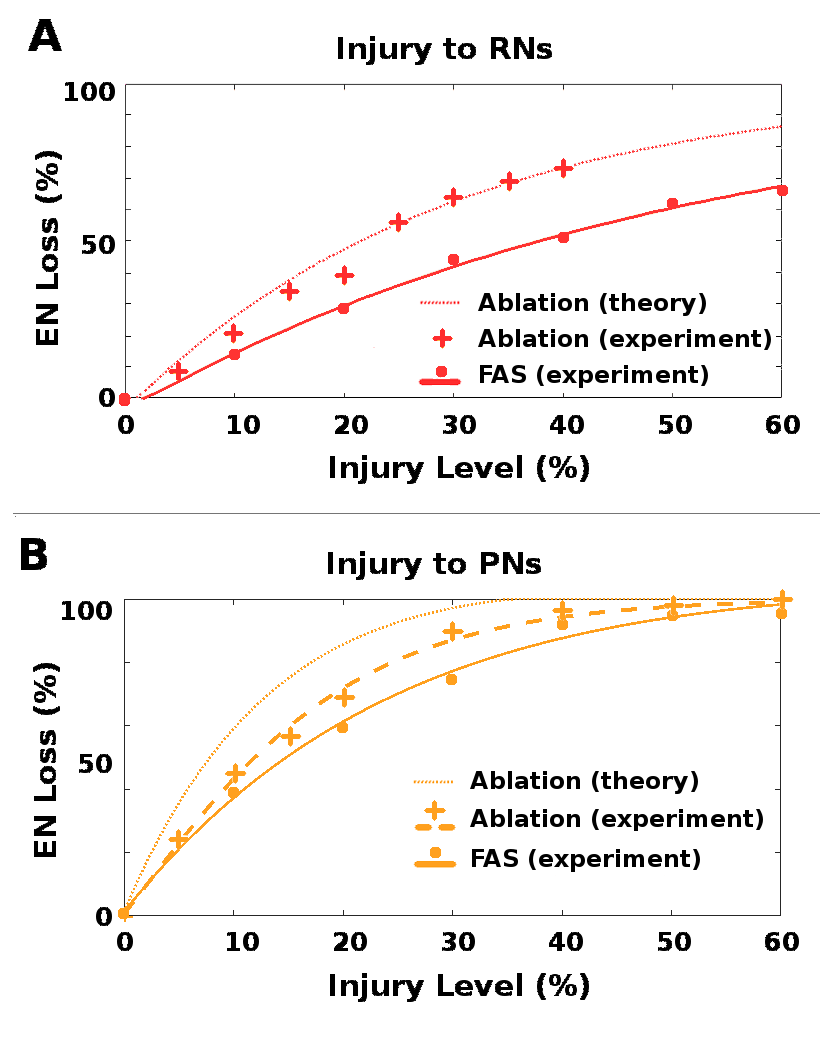}  
\caption{\small { { \bf Level of Ablation and FAS injury needed to induce a given EN loss:} 
The relationship of ablation and FAS injury effects varied with injury location, implying that ablation is an unreliable proxy for FAS injury. 
In both plots: dots and crosses are experimental results for FAS and ablation respectively.
Solid and dashed curves are fits to experimental results for FAS and ablation respectively.
The dotted curve is the theoretical result for ablation, given the FAS experimental results and a 1.75$\times$ ratio (see Section \ref{sectionRatioCalculation}).
{\bf{A:}} When RNs were injured, ablation induced loss to EN response consistent with theory (1 unit Ablation $\approx$1.75 units FAS injury). 
Curves do not meet at origin because they are fits to data points.
{\bf{B:}} When PNs were injured, ablation induced a much smaller loss than theory (1 unit Ablation $\approx$1.25 units FAS injury).
This can be seen by looking  at horizontal lines (\textit{i.e.} fixed EN loss), and comparing injury levels that induce this loss for (from right to left) experimental  FAS injury, experimental ablation, and theoretically-expected ablation.
A large gap exists between experimental and theoretical ablation levels required to induce a given EN loss.
  } 
  }
\label{pieVsAblationResultsPanel}
\end{center}
\end{figure}

The qualitative effects of ablation and FAS were similar, at each injury site.
However, the relative quantitative effects of the two injury types varied greatly depending on the site of injury. 
For RN channel damage, ablation effects were roughly in line with that predicted by theory for large homogeneous populations, \textit{i.e.} 1.75$\times$ FAS damage. 
This match between theoretical and experimental ablation effects is seen in Fig.\ref{pieVsAblationResultsPanel} A.
The match makes sense given the assumptions on number and distribution of RNs stated above.
 
In contrast, ablation injury to the PN channel was much less harmful relative to FAS-like injury than predicted by theory.  
For example, 10\% ablation would theoretically  induce the same EN loss as 17.5\% FAS injury and 20\% ablation would correspond to 35\% FAS injury.  
However, in our experiments 10\% ablation corresponded to only $\sim$12\% FAS injury (a ratio of 1.2), and 20\% ablation corresponded to only $\sim$25\% (a ratio of 1.25).
This effect is seen in Fig.\ref{pieVsAblationResultsPanel} B by following horizontal lines, which correspond to equivalent EN loss, and comparing the percentage injury levels of (from right to left) experiment FAS, experiment ablation, and theoretical ablation.
The experimental ablation levels required  to induce a fixed EN loss were much closer to the FAS levels than theory predicted (ratio $\sim$1.25 instead of 1.75).
 
We remark that this measured discrepancy between theoretical and actual effects is not at the site of injury, but at the downstream (readout) neurons, \textit{i.e.} after the impact of the upstream injury has been nonlinearly modulated by moving through the cascaded system.

This variability in the ratio of ablation injury  to equivalent FAS injury, dependent on which neurons are injured, suggests that ablation may be an unreliable proxy for naturalistic neuronal damage in some contexts.

\section{Discussion}

Our simulations indicate that the neural mechanisms and motifs we tested have clear injury-mitigation properties. 
In this section we suggest mechanisms by which these structures might protect readout neuronal activity from upstream injury.
We note that from a functional point of view, overall resilience of a cascaded system depends partly on whether upstream 
regions can avoid damage, but mainly on whether downstream units can still transmit key readout signals to the rest of the 
body despite upstream damage. 
Cascaded networks are ubiquitous among biological neural systems, so the principles discussed in the moth's olfactory network may be applicable to other settings. 
We also discuss the discrepancy between axonal swelling injuries \textit{vs} ablation injuries. 
Finally, we argue that robustness to injury is a key principle of biological neural design.

 
\subsection{Hebbian plasticity-induced recovery from injury } 
Learning in the moth olfactory network occurs via a combination of octopamine stimulation and Hebbian growth. 
Octopamine stimulation temporarily boosts neural firing rates during reinforcement by sugar reward, while Hebbian updates strengthen the synaptic weight $w_{ab}$, between two neurons $a$ and $b$,  proportionally to the product of their firing rates:  
\begin{equation}
\Delta w_{ab} \propto f_a(t) f_b(t).  
\end{equation}
Injuries to the upstream regions of a network result in spike deletions and weaker encodings arriving at downstream neurons. 
If the damaged region cannot activate downstream neurons with the existing synaptic connection strengths, there is a functional loss of information. 
However, the combination of octopamine stimulation and a Hebbian update mechanism can evidently mitigate or reverse this effect.
We note that the original injured neurons are not themselves repaired, as plasticity only boosts downstream synaptic connections.

We note that the Hebbian mechanism alone is not sufficient to repair damage, because updates to synaptic strengths require sufficiently strong FRs in both the incoming and the receiving neurons. 
If injury reduces the FRs of upstream neurons, and these reduced inputs lead to lower FRs in downstream neurons, Hebbian updates are significantly degraded.
Thus, octopamine-induced stimulation of the injured upstream neurons is crucial to post-injury plasticity, because it temporarily boosts the FRs of injured upstream neurons back to levels that enable the Hebbian mechanism to strengthen the relevant connections.

\begin{figure*}[t]
\begin{center}
\includegraphics[width = .95 \textwidth] {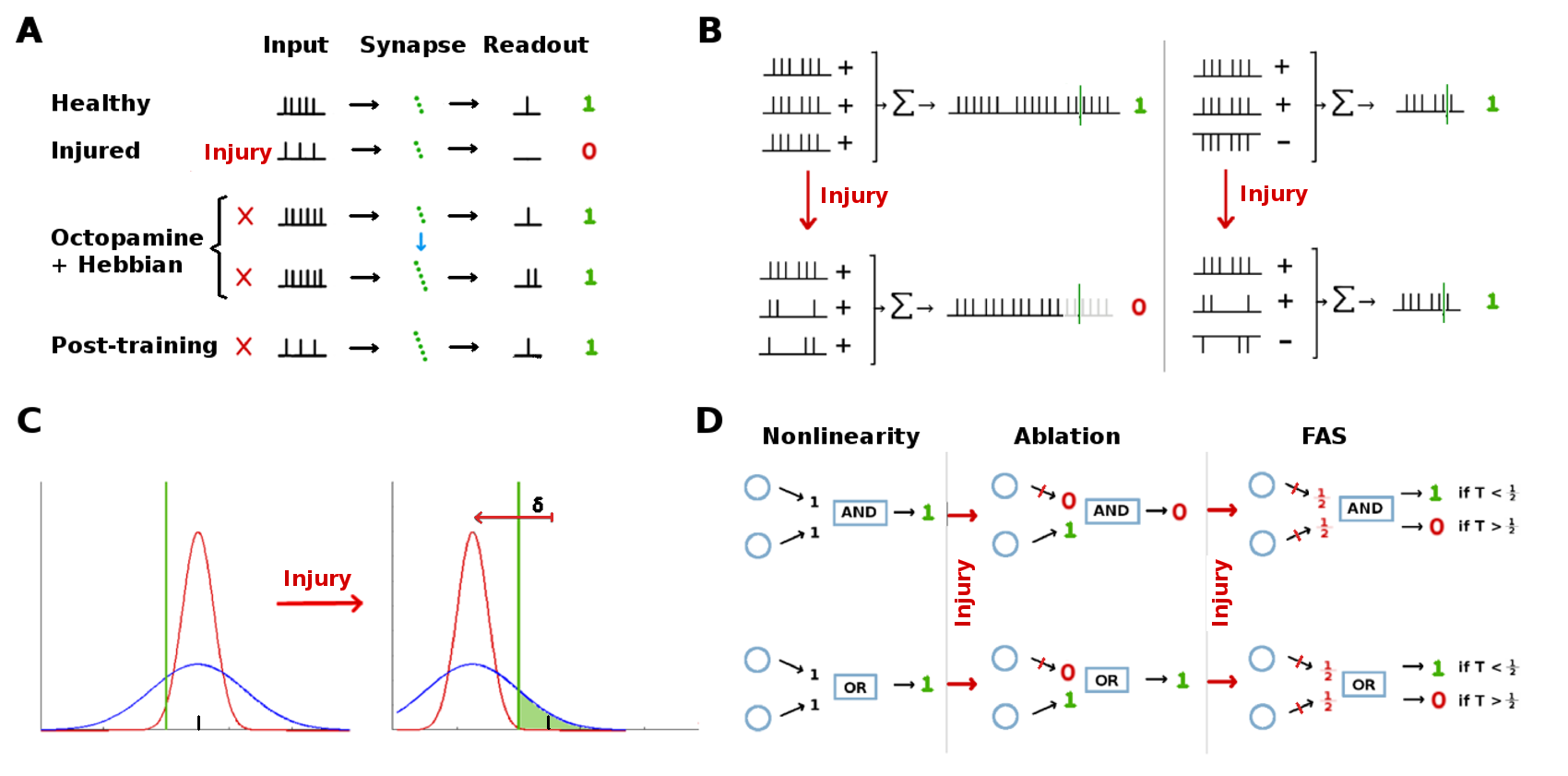}   
\caption{ \small {  {\bf Injury mitigation hypotheses:} 
In a cascaded network, various architectures can mitigate the effects of injury to upstream 
neurons by protecting or restoring functionality of downstream units. 
{\bf{A:}} Learning itself can compensate for injury: 
Octopamine temporarily stimulates the damaged neuron, allowing 
Hebbian growth to strengthen downstream synaptic connections.Though the injured neuron's 
signal is not restored, the downstream neurons receive an amplified input, cancelling out the 
injury.
{\bf{B:}} Parallel inhibitory channels can reduce the effect of generalized injury by spreading 
damage among excitatory and inhibitory signals, so that losses cancel out in terms of inputs 
to downstream neurons.
{\bf{C:}} Wide noise envelopes on upstream neuron outputs can protect the strongest stimulus 
responses from injury-induced attenuation $\delta$, to the degree that their std dev $\sigma > \delta$. 
This allows the injured neuron's strongest responses to still exceed their activation threshold 
(green line) for downstream neurons, protecting downstream functionality.
{\bf{D:}} Two simple examples of non-linearities that can result in qualititative change in the relative 
effects of ablation and FAS-like injury: 
In an AND gate, ablation can have a worse effect than FAS downstream, 
depending on the gate's input threshold $T$. In an OR gate, ablation can be harmless, while FAS 
can have a worse effect downstream, depending on $T$.
}  }
\label{hypothesesPanel}
\end{center}
\end{figure*}

\noindent We propose that degraded firing rates in downstream neurons are restored via the following mechanism (see schematic 
in Fig.\ref{hypothesesPanel}A):
\begin{enumerate}
\item Octopamine temporarily increases the firing rates of injured upstream neurons.
\item The transient boosted encodings are sufficient to trigger firing in the downstream neurons
with the existing synaptic connection strengths.
\item Since neurons on both sides of the plastic connections are firing, Hebbian growth  
strengthens their connections.
\item Firing rates from the injured upstream region return to their reduced rate once octopamine is withdrawn. 
However, due to the stronger synaptic connections, these encodings are now sufficient to trigger the downstream neurons. 
This restores the transmission of key information to the rest of the body.
\end{enumerate} 

Because the \{octopamine stimulation + Hebbian updates\} learning mechanism is automated, \textit{i.e.} hard-wired as a reward mechanism for adaptive stimuli such as sugar, it acts as a passive injury mitigation system (absent injury, it serves to boost network responses to adaptive stimuli).
Since learning is activated repeatedly throughout life (by any rewarding stimulus), it can be expected to act post-injury as an automatic repair mechanism.
Alternately, learning can be viewed as a built-in tuning mechanism that in event of injury serves to restore network responses towards their pre-injury states. 

We hesitate to call this learning mechanism homeostatic, even though in the context of injury it automatically moves the system towards a prior state, because the restoration is one-way. 
Learning will not  revert responses that have been previously strengthened (by learning itself).  
Rather, it is an automated mechanism for tuning a network towards stronger responses to adaptive stimuli, which in the event of injury has a homeostatic effect.  


\subsection{Inhibitory neurons and protective canceling out effect}
The moth olfactory network has both excitatory projection neurons and inhibitory projection neurons that feed-forward from the antennal lobe to the mushroom body. 
We propose a mechanism to explain how this can protect downstream neurons from the effects of upstream damage, assuming downstream dynamics depend on the summed input from upstream neurons:

\begin{equation}
(\bf{w}\cdot\bf{u}) = \bf{w^+}\cdot \bf{u^+}  - \bf{w^-} \cdot \bf{u^-} \text{, where}
\end{equation} 
\begin{eqnarray*}
{\bf{w^+}} &=& \text{connection weights from excitatory neurons}\\  
{\bf{u^+}} &= & \text{FRs from upstream excitatory neurons} \\
{\bf{w^-}} &= & \text{connection weights from inhibitory neurons} \\
{\bf{u^-}} &= & \text{FRs from upstream inhibitory neurons.}  
\end{eqnarray*}
When FAS-like injury is applied to the PN+QN  pipeline in our neural architecture (mimicking the outcome of a physical shock), 
the net effect on the summed signal reaching downstream target neurons varies 
according to the proportion of QNs to PNs (${\bf{u^-:u^+}}$, assuming uniform weights $\bf{w}$).
When all feed-forward signals are excitatory (\textit{i.e.}, ${\bf{u^-}}$ = 0), injury will always reduce the summed input reaching a downstream neuron. 
If QNs exist, however, and both PNs and QNs share the same exposure to injury, then the overall reduction to the summed input will be mitigated on average, since any injury to QNs will  increase the summed input, offsetting decreases due to PN injury. 
A schematic of this ``cancelling out'' mechanism is shown in Fig.\ref{hypothesesPanel} B. 

The injury resistance provided by high QN counts comes at a cost to other functionalities, \textit{e.g.}  higher spontaneous EN noise relative to odor response. 
Presumably, biological networks have QN counts which optimally balance the benefits of injury mitigation on one hand versus the need for high signal-noise-ratio, as well as other concerns such as the energy cost to the organism. 
If  the QN counts are low (\textit{e.g.} QN:PN $\leq$ 20\%, as in the moth), this injury mitigation benefit is likely less important relative to other  architectural or functional constraints.
We note that learning and plasticity are not pre-requisites for this mechanism. 


\subsection{Upstream noise protects downstream behavior} 
Suppose that the behavioral response is preserved after injury if at least a subset of stimuli elicit downstream responses that exceed action-triggering thresholds. 
In this case, a large noise envelope on upstream neurons may help protect the network's functionality. 
 
Assume the firing rate of an upstream neuron FR responds to stimuli following a Gaussian distribution $N(\mu, \sigma)$, and that it needs to exceed a threshold $T$ to activate downstream neurons. 
If the neural damage reduces this FR in average by $\delta$, a large noise envelope (large $\sigma$) will ensure that some post-injury responses still exceed threshold, \textit{i.e.} that $\mu  - \delta + \sigma \geq T$.
This idea is sketched in Fig.\ref{hypothesesPanel}C for two FRs  characterized by $N(\mu , \sigma_1 )$ and $N(\mu ,\sigma_2 )$ with $\sigma_1 > \sigma_2$.
 
Our experiments indicate that AL noise does enable the highest EN responses to exceed threshold after injury, even as the average EN response drops.
However,  the injury mitigating benefit of increased upstream noise comes at a cost to other system functionalities, \textit{e.g.} it reduces signal-to-noise ratio.  
Noise levels in biological networks (such as in the antennal lobe) may represent an evolved/optimized trade-off between injury mitigation effects and negative side-effects such as reduced SNR. 
We note that the sparsity of the MB acts as a powerful noise filter \cite{delahuntMoth1}. 
Plasticity is not a pre-requisite to this mechanism.

\subsection{Ablation is a poor proxy to biological injury}
Neuronal injuries are often modeled in a binary way, \textit{i.e.} by treating a neuron and/or its connections as either fully functional or fully impaired. Our results indicate, however, that in some situations ablations are a poor proxy for more naturalistic FAS-types of 
injuries regarding effects measured downstream from the injury site.   

When the neuron population to be injured is large, and has pooled outputs to the next layers (in our model, the antennae/RNs), ablation maps to FAS injury in a predictable manner due to averaging effects over the population (see section \ref{sectionRatioCalculation}).
However, when the neuron population is small (in our model, the PNs) the effects of ablation \textit{vs} FAS are not predictable.
Ablation of PNs had much lower impact than large-population theory would predict

Our key finding is that ablation effects are inconsistent relative to FAS-like effects, depending on the location and characteristics of the injured neurons.
This calls into question the value of ablation as a proxy for naturalistic neural injuries.
We suggest that  in systems with large numbers of somewhat interchangeable units (\textit{e.g.} the 30,000 RNs) ablation is a suitable way to model injury; while in systems with smaller numbers of specialized neural units (\textit{e.g.} the AL with 60 glomeruli) ablation is a poor injury model.
This unsuitability is sharpened when the actionable effects of injury are measured downstream from the regions injured, because there is a complex interplay between the injuries and network nonlinearities, making the outcome somewhat unpredictable. 
Simple examples of possible effects of non-linearities (AND and OR gates) that might cause ablation vs. FAS-like injury models to diverge are shown in Fig.\ref{hypothesesPanel} D.

\subsection{Limitations} 
Our computational model assumed only one readout neuron and one broadly-activating odor. 
A more realistic assessment  of injury and mitigation might involve several readout neurons to allow for disparate effects on various readouts, and might use more narrowly-activating odors. 
We do not know if the PN+QN  channel is a realistic target for injury.
We chose it in order to investigate deficits caused by injury to innermost hubs. 
Our study certainly did not exhaust all potentially interesting combinations of structures-under-test and injuries.
 
 
\section{Conclusion}

We  investigated the moth olfactory network with the goal of understanding how its basic architectural components serve to make sensory processing robust to injury.   
Since most organisms are exposed to neuronal damage throughout their lives, it is important to understand how such neuronal circuits are structured to maintain functionality despite impairments. 
In this work, we showed explicitly how certain structural and functional network motifs act as injury mitigation mechanisms.
 Specifically, we examined how \textit{(i)} Hebbian learning, \textit{(ii)} high levels of noise, and \textit{(iii)} presence of both inhibitory and excitatory connections, can support overall robustness to injury in the olfactory system in the \textit{Manduca sexta} moth.  

Our findings indicate that, in addition to accurate sensory processing, biological neural networks such as those found in the moth olfactory system hold robustness to injury as a central design principle.  
Our findings also suggest an additional hypothesis:   
Plasticity coupled with neuromodulatory stimulation, now central to learning, may have originally evolved as a repair mechanism for neural systems to offset injury and maintain function, and was only later ported to the task of developing responses to new information (exaptation). 
If this is the case, then the gift of learning is due originally to the exigencies of brain damage.

Our results also show that these architectures can in fact cause worse performance by some other performance metrics, \textit{e.g.} SNR. 
Thus, trying to explain these architectures from the point-of-view of, for example, information theory risks running against the fact that they are actually suboptimal according to that particular lens.  
A more comprehensive and nuanced framing of the neural signal processing task, positing multiple design goals including injury mitigation, can enable better understanding of neurosensory processing.  
That is, a neural architecture can be understood only if its injury mitigation function, and the trade-offs between this and other desired functions, are considered.
Indeed, it is possible that some neural structures and mechanisms, including the ability to learn, are best understood as evolutionary solutions to the challenge of maintaining function despite injury.


\section{Materials and Methods} 
In this section, we first detail the computational model \textit{MothNet} used in all experiments.
We then describe focal axonal swelling (FAS), a characteristic form of neuronal injury utilized as a model of damage, and how it was applied to the network. 
Lastly, we provide details about the experimental setups involved in our key findings.
\subsection{\textit{MothNet} architecture} \label{sectionArchitecture}
\noindent We use close variants of the model of the moth olfactory network (\textit{MothNet}) developed in \cite{delahuntMoth1}, modifying the architecture features as needed for each experiment.
We provide only selected details about the architecture here.
For a fuller description, please see \cite{delahuntMoth1} and its associated codebase, as well as the Matlab codebase for \textit{MothNet} and injury simulations available at \cite{delahuntMothInjuryCodebase}. 

\textit{MothNet} uses firing rate dynamics for neural firing rates \cite{dayan2001}, evolved as stochastic differential equations \cite{higham2001}, and Hebbian plasticity for synaptic weight updates \cite{hebb}.
A table of governing equations and parameters are given below.  

\noindent For the experiments in this work, the relevant architectural structures of the MON  were: \\

\noindent \textit{Antennae:}  ~\\
Network structures in which chemical receptors detect odor and send signals to the Antennal Lobe via receptor neurons (RNs). 
There are approximately 30k RNs (assumed here to be $\approx$ 500 per glomerulus) that \textit{MothNet} combines into one averaged RN per glomerulus. 
All injury protocols applied to RNs accounted for this many-to-one abstraction (for details, see  Section \ref{sectionInjuryMethods}. 
While RNs can respond to several atomic odors, \textit{MothNet} makes the simplifying assumption that each RN responds to exactly one, and that all RNs responsive to a certain target innervate the same glomerulus in the AL. 
Thus atomic odors and AL glomeruli are assumed to correspond 1-to-1.\\
 
\noindent\textit{Antennal Lobe (AL):} ~\\
The AL acts as a pre-amp, converting the weak electrical signals into an output signal strong enough to tolerate noise and allow further processing. 
It also modulates the odor's encoding via intra-AL lateral inhibition. 
The AL structure contains approximately 60 neural units (glomeruli) which process odors and send excitatory signals (via projection neurons  PNs, $\approx$ 5 per glomerulus) and inhibitory signals (via QNs) downstream to the mushroom body. 
In the version of \textit{MothNet} used here, QNs have dendrites in one glomerulus, rather than in several (as in the actual MON), and were parametrized as inhibitory analogs to PNs, with the same connection distributions and FR behavior as PNs.
This enabled us to vary the ratio of QNs to PNs from 0 to 1.4 according to experiment. \\

\noindent\textit{Mushroom Body (MB):} ~\\
The MB is a high-dimensional ($\approx$ 4000 neurons), sparsely-firing structure that encodes odor signatures and memories, and contains plastic synaptic connections.
Odor responses feed-forward from the AL to the MB via PNs (excitatory) and (in moths) QNs (inhibitory).
MB neurons feed-forward to the Readout Neurons. 

Sparsity in the insect MB is enforced by global inhibition either from  the Lateral Horn (moth) or from the MB itself (\cite{bazhenovStopfer2010}, \textit{drosophila} \cite{lin2014}, locust \cite{papadopoulou2011,guptaStopfer2012}). 
In \textit{MothNet}, MB sparsity is enforced by injecting a time-varying inhibitive input term into all MB neurons,  such that only the most strongly-excited neurons  fire and the percentage of MB neurons with positive FRs is fixed (at \textit{e.g.} 5 to 15\%).
This inhibitive term models input from the Lateral Horn, which is thus implicitly, not explicitly, modeled by global inhibition  of the MB.\\

\noindent\textit{Readout Neurons (ENs):}~\\
Odor codes in the Mushroom Body (MB)  feed-forward  to Readout 
Neurons (Extrinsic Neurons, ENs), which are assumed to act as decision-making neurons. Strong EN responses 
trigger actionable messages (such as ``fly upwind'').  
When assessing effects of injury, we focus on the ENs, since these represent the final, actionable output of the system. 
\textit{MothNet} posits one EN, whose output firing rate serves to measure the functional effects of upstream injury.\\

\noindent\textit{Octopamine effects:}~\\
Octopamine is crucial to learning in the MON. 
In MothNet, effects of octopamine instantiate in three ways: 
\textit{(1)} It stimulates AL neurons (RNs, LNs, PNs), making them more responsive to excitatory inputs and less responsive to inhibitory inputs (see Eqns in Section \ref{sectionGoverningEqns}); \textit(2) It activates plasticity in AL$\rightarrow$MB and MB$\rightarrow$EN synaptic connections (in MothNet this simply means that plasticity in these regions is ``turned on'' when octopamine is non-zero; \textit(3) It reduces the sparsity parameter of the KCs in the MB, to model that it makes KCs less responsive to Lateral Horn inhibition.
Our experiments indicated that items \textit{1} and \textit{2} are necessary for learning (as in  \cite{hammer1995,hammerMenzel1998}), while item \textit{3} is optional \cite{delahuntMoth1}. 
%

\subsection{Moth template parameters} \label{injuryMothTemplates} 
%
%
In each experiment, over 30 moth instances (per data point) were randomly generated from the \textit{MothNet} template defining the architecture, which included biologically-plausible choices for 
numbers of neurons, synaptic connection weights,  how odor is projected onto the glomeruli of the AL, as well as learning rates and SDE time and noise constants.
To generate connection matrices, non-zero connections were randomly assigned according to architectural constraints and template parameters, then non-zero connection weights were then drawn from gaussian distributions with parameters dependent on types of neuron.
For full details of how \textit{MothNet} instances are generated from templates, please see \cite{delahuntMoth1}.
The templates were realistic in the senses of having \textit{(i)} PN firing rate behavior matching \textit{in vivo} data from live moths and \textit{(ii)} architecture parameters that match what is known from the literature \cite{delahuntMoth1}.
Some templates were moved to the boundaries of, or out of, a known realistic regime by varying key parameters-under-test when required by the experiment.

\begin{enumerate}
\item The number of QNs per glomerulus varied from 0 to 7, in order to test the injury-mitigating effect of inhibitory QNs in parallel with PNs in the feed-forward AL $\rightarrow$MB channel. PNs were fixed at 5 per glomerulus, as in moths. 
In moths, each QN has dendrites in several glomeruli. 
In our experiments, each \textit{MothNet} QN has dendrites in exactly one glomerulus (like PNs), to make QN:PN ratios meaningful. 
Moths may have QN:PN ratio  $\approx$ 0.2 (\textit{i.e.} relatively few QNs), insofar as a ratio can be estimated (actual values are not known).
We note that more QNs means that more individual MB neurons receive extra inhibition.
The global inhibition mechanism modulated by the Lateral Horn is distinct, and is not affected by QN numbers. \\
\item The level of Gaussian noise affecting all AL neurons varied from 0 to 1.33, where 
1.0 represents natural levels (\textit{i.e.} fitted to \textit{in vivo} data). The purpose of this noise 
range was to test for any injury-mitigating values for the AL structure. \\
\item For experiments examining effects of learning and of AL noise, we set the number of QNs per 
glomerulus to zero. Setting the number of QNs equal to 2 (equivalently, QN:PN ratio = 0.4) gave similar 
results. These QN values remain close to that of realistic models.
\end{enumerate}

\noindent The extremes of the parameter regimes described above deviated significantly from calibrated models, 
and sometimes created moths with untenably noisy, dysfunctional EN responses to odor.  
Thus, we discarded moths with naive EN odor response-to-spontaneous FR (SSNR)
outside an envelope defined by $\frac{\mu(F)}{\mu(s)}$ $<$ 12. 
These comprised about 12\% of moths generated, with the percentage depending on the varied parameters: 
Templates with high numbers of QNs and/or very high AL noise had more rejected moths; templates with few QNs and normal/low AL noise had few rejected moths. 
Extra moths were generated as needed to match numbers across all experiments, so that each  \{parameter-under-test, injury level\} pair (\textit{e.g.} ``4 QNs, 50\% FAS injury'') had $\geq$ 30 moth instances. 

\subsection{FAS-like injury}

Focal Axonal Swellings (FAS) is a neural injury associated with traumatic brain injury (TBI), typically caused by physical shock. 
Examples in current events include blast injuries from recent wars, as well as impact injuries in contact sports. 
FAS presents as swollen neural axons (the signal delivery pipelines) with dramatic diameter changes,  
causing signals from the upstream source to be diminished or lost entirely before reaching downstream 
target neurons \cite{maiaCompromisedAxonal}.
This degradation can be expressed as reduced FRs from upstream neurons, characterized in a computational model by \cite{maiaReactionTime}, which found that signals traveling down an injured axon are attenuated to greater or lesser degree according to the amount of swelling and the firing rate of the signal.
While ablation is a ready and oft-used means to model neural injury, it imposes a binary “all-or-nothing” effect which is not present in FAS injuries. 
In these experiments we model neural injury according to \cite{maiaReactionTime}, hereafter “FAS type” or “FAS”.

\subsubsection{FAS in the FR model context}
While FAS models effects at the level of spike trains, it also has a meaningful representation in Firing Rate models such as \textit{MothNet}.
In particular, unlike ablation, FAS causes reduced but still non-zero FRs. 
In addition, the low-pass filtering effect of FAS, which impacts closely-bunched clusters of spikes more than sparse spikes, in analogous manner impacts high FRs more strongly than low FRs. 
Thus FAS, applied in a FR model, results in neuron FRs being reduced but not ablated, with high-FR neurons affected more strongly than low-FR neurons.
For a fixed amount of total damage, FAS results in relatively many partially-damaged neurons, while ablation results in relatively few fully-destroyed neurons.
The structure of FAS injuries allows a principled conversion to the FR-based model context, detailed below.


\subsubsection{Injury methods} \label{sectionInjuryMethods}

FAS due to physical trauma does not affect all neurons in a targeted brain region equally, nor does it operate in an 
``all or nothing'' way \cite{wang}. We used an injury regime derived in \cite{maiaReactionTime}, which calculates the 
fractions of injured neurons falling into each of four injury types: FR unaffected (transmission); FR cut by half (reflection), 
FR destroyed (ablation), or FR filtered according to
\begin{equation}
f_{injured}(s) = F(f_{healthy}(s) ),
\end{equation} 
where $f_*$ = firing rate, $s$ is a stimulus, and $F$ is a lowpass filter.
FAS injury fractions were as follows: 15\% transmission, 35\% reflection, 35\% ablation, and 15\% low-pass filtering, where the low-pass filtering generally multiplied FRs by a factor of 0.7 to 0.9 depending on initial FR (taken from \cite{maiaCompromisedAxonal}). 
Neurons in the target group were randomly selected for injury according to the percentage specified in the particular experiment (0 to 60\%). 
Each injured neuron in the target group was then randomly assigned one of the damage types. 

Applying this injury regime to populations of PNs and QNs in the model is straightforward, since these neurons are 
modeled one-to-one (\textit{i.e.}, one neuron in the model represents one actual neuron).
For example, given 300 PNs and a FAS injury level of 50\%, 150 PNs would be randomly assigned (in fixed proportions) to have their FRs multiplied by either 1 (transmit), 0.5 (reflect), 0 (ablate), or variable (low-pass filter).
At an ablation level of 50\%, 150 PNs would be randomly  assigned to have their FRs zeroed out.

Injury to RNs was handled differently because in \textit{MothNet} each RN (inputting to one glomerulus in the AL) stands for $\approx$ 500 RNs in the moth.
Injury of RNs was handled as follows, leveraging averages over large numbers of homogeneous neurons:

\begin{enumerate}
\item The FAS injury level $m$ (\% of neurons injured) was converted into a theoretically equivalent ablation injury level $n$ (\% of neurons ablated), where $m = 1.75n$ (see calculation below).  
\item Each RN's FR was multiplied by ($1-n$) (\textit{i.e.} attenuation), since the glomerulus had fewer inputs. 
\item The RN noise parameter was multiplied by $\frac{1}{\sqrt{1-n}}$, since the glomerulus was averaging fewer inputs, giving  less noise reduction from averaging.
\item 15\% of FAS injury was low-pass filtering, which depends on the injured neuron's firing rate. 
This part of the FAS injury to RNs varied according as the RN firing rate varied.
So the actual $n(t)$ affecting a given RN fluctuated slightly with odor inputs.
\end{enumerate}

\subsubsection{Theoretical correspondence of FAS to ablation} \label{sectionRatioCalculation}
We wish to compare levels of ablation and FAS injury, using as a measure the percentage of neurons injured.
Since FAS does not usually destroy a neuron's FR while ablation does, we expect that ablating $n$\% of neurons will cause the same average drop in summed population FR as FAS injury to $m$\% of neurons where $m> n$.

We assume a large homogeneous population of neurons.
Given a fixed percentage of FAS injury, we can estimate  the ablation level that gives the same average total reduction to summed FR, as follows:
\begin{align*}
& \text{FAS applied to 100 units} \\ 
&\implies 15~ \text{transmit} + 35~ \text{reflect}+ 35~\text{ablate} + 15~\text{lowPass}  \\
&\implies (15 \times 1) + (35 \times 0.5) + (35 \times 0) + (15 \times 0.7) \approx 43 \\
&\implies \text{ablation applied to 57 units}\\
&\implies \text{conversion rate } = 100 / 57 \approx 1.75.
\end{align*}  
\textit{e.g.} 20\% ablation nominally corresponds to 35\% FAS injury, in terms of total reduction in summed FRs over the injured population.  \\

\subsubsection{Location of injured regions}

Two sites were targeted independently for FAS injury (RNs or PNs+QNs), with the choice determined by the experiment.
See Fig.\ref{injuryRnsPns}. \\

\noindent\textit{RNs:} ~\\
We posited the Antennae $\rightarrow$ AL channel, \textit{i.e.} the RNs, as a likely site for FAS 
injury due to their exposure to external impacts. Fig.\ref{injuryRnsPns} (red stars) shows this injury
location.
There are about 30k RNs, which we divide as 500 RNs responding to each of 60 atomic odors. 
These atomic-focused groups send their inputs to a single glomerulus in the AL, where their inputs are averaged to reduce noise. 
The receptors for a given glomerulus are distributed across the antennae. 
Thus, we expect injury to an antenna to affect each glomerulus' RN input roughly equally, and to affect a roughly equal percentage of each glomerulus' 500 inputs. \\

\noindent \textit{PNs+QNs:} ~\\
There is a channel that carries PN (and QN) axons from the AL to the MB. 
We modeled damage to this channel by injuring both PNs and QNs with equal probability.
Fig.\ref{injuryRnsPns} (orange stars) shows this injury location. We assumed that the  degree of sparsity enforced on the MB (either by the Lateral Horn, or by global self-inhibition by the MB) remained stable over modulations of input signal from the AL.

\subsection{Simulation protocols} \label{simulationProtocols}

Each experiment consisted of many moth instances, all generated from the same template with only the parameters-under-test varied.
Over 30 moth instances (trials) were run for each  parameter pair (\textit{e.g.}``4 QNs, 50\% FAS injury") with training on 5 odor puffs.  
Odors were randomly generated and projected onto the AL broadly ($\approx$ 25 glomeruli targeted).
A single trial consisted of first injuring and then training a single moth, in five or six stages:  

\begin{enumerate} [start=0]   
\item (for discrimination experiments only) Pre-training to ensure initial discrimination: Trained odor plus octopamine were applied (3 odor puffs) with Hebbian plasticity activated.
\item Pre-injury baseline: Odor was applied without octopamine (15 odor puffs, each 0.2 mSec)  to assess naive EN odor response. 
\item Injury was applied.  
\item Post-injury odor: Odor was applied without octopamine (15 odor puffs)  to assess the effects of injury on EN odor response. 
\item Training: Odor plus octopamine were applied (5 to 15 odor puffs), with Hebbian plasticity activated,  to train the system (plasticity was coincident with octopamine, details in section \ref{sectionArchitecture}). 
\item Post-Training odor: Odor was applied without octopamine (15 odor puffs)  to assess post-training EN odor response.
\end{enumerate} 

\noindent Firing rate $f(t)$ from a single EN was recorded, to track the actionable effect of injury and training on the system.
A timecourse of EN firing rates, from a typical experiment, is shown in Fig.\ref{typicalEnTimecourse}. 

\subsection{Additional simulation details} \label{detailsByExperiment}

\noindent \textit{Learning  compensates  for injury:} ~\\
Moth templates were biologically plausible, in the sense that their AL behavior matched \textit{in vivo} data, and they demonstrated learning behavior.
Moths in this experiment had 0 QNs, \textit{i.e.} no feed-forward inhibitory signals from AL$\rightarrow$MB (2 QNs per glomerulus gave very similar results).
In separate experiments, either the RN channel (Antennae$\rightarrow$AL) or the PN+QN  channel (AL$\rightarrow$MB) was injured with FAS (0\% to 60\%).\\

\noindent\textit{Parallel inhibitory neurons protect EN responses:} ~\\
Moths were generated from biologically plausible template but with \{0, 2, 4, 5, or 7\} QNs per 5 PNs. The corresponding QN:PN ratios are \{0, 0.4, 0.8, 1.0, or 1.4\}.  
AL noise  was set to a natural level (matching \textit{in vivo} data). 
The PN+QN  channel was injured with FAS (0\% to 60\%). PNs and QNs were treated equally in terms of injury. \\

\noindent\textit{AL noise preserves the highest EN responses:} ~\\
Moths templates had AL noise level from 0 to 1.33, where 1.0 corresponds to natural AL noise. 
Moths in this experiment had only excitatory PNs (\textit{i.e.} \#QNs = 0). Moth templates with 2 QNs per glomerulus gave similar results. 
The RN channel was injured with FAS from 0\% to 60\%. \\

\noindent\textit{Comparison of ablation \textit{vs} FAS injuries:}~\\  
To compare effects of ablation \textit{vs} FAS injury, we ran parallel experiments using ablation and FAS.
Injury levels were set between 0\% to 60\%, using either FAS type or ablation.
FAS injury fractions were 15\% transmission, 35\% reflection, 35\% ablation, and 15\% low-pass filtering ($\sim 0.7x - 0.9$x in most cases)  \cite{maiaReactionTime}. \\

\begin{figure*}[t]
\begin{center}
 \includegraphics  [width= .65\textwidth]{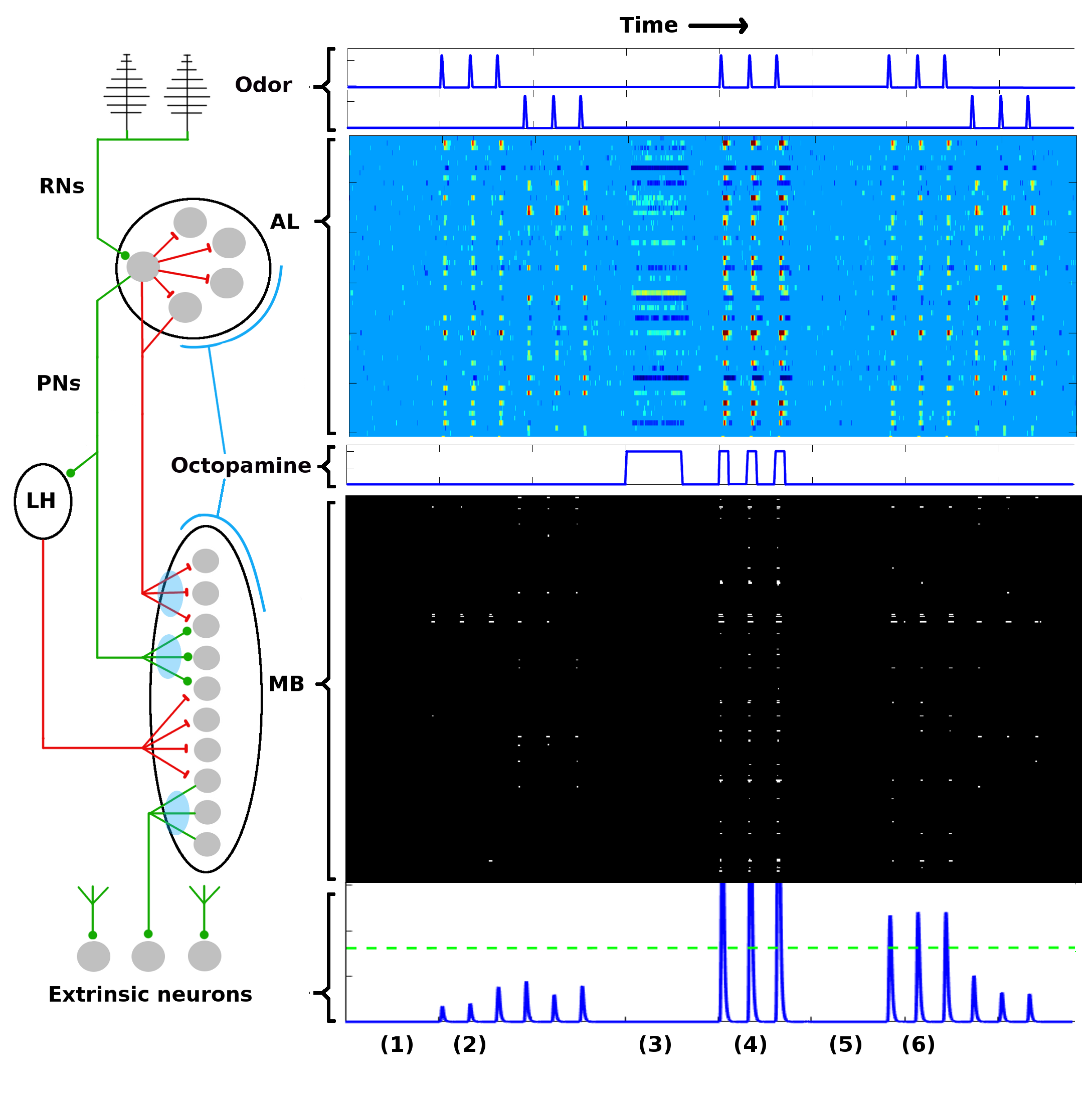}  
\caption  {\small{ {\bf Overview of AL-MB.}
Left: System schematic. Green lines show excitatory connections, red lines show inhibitory connections. 
Lateral horn inhibition of the MB is global on the MB. 
Light blue ovals show plastic connections into and out of the MB.\newline
Right: Neuron outputs for each network (typical simulation) with time axes aligned. 
Timecourses are aligned with their regions in the schematic. 
The AL timecourse shows all responses within $\pm$ 2.5 std dev of mean spontaneous rate as medium blue. 
Responses outside this envelope are yellow-red (excited) or dark blue (inhibited). 
MB responses are shown as binary (active/silent). 
Timecourse events are as follows: 
1. A period of no stimulus. All regions are silent.
2. Two odor stimuli are delivered, 3 puffs each. AL, MB, and ENs display odor-specific responses. 
3. A period of control octopamine, \textit{i.e.} without odor or Hebbian training. 
AL response is varied, MB and EN are silent.
4. The system is  trained (octopamine injected) on the first odor. 
All regions respond strongly.
5. A period of no stimulus. 
All regions are silent, as in (1).
6. The stimuli are re-applied. 
The AL returns to its pre-trained activity since it is not plastic. 
In contrast, the MB and EN are now more responsive to the trained odor, while response to the untrained odor is unchanged. 
Green dotted line in the EN represents a hypothetical ``action" threshold. 
The moth has learned to respond to the trained odor.
 } }
\end{center}
\end{figure*}

\noindent \textit{Plasticity:}~\\
Our model assumes a Hebbian mechanism for growth in synaptic connection weights \cite{hebb,cassenaer,roelfsema2018}. 
The synaptic weight $w_{ab}$ between two neurons $a$ and $b$ increases proportionally to the product of their firing rates 
(``fire together, wire together''): 
 
$\Delta w_{ab}(t) \propto f_a(t) f_b(t). $ 

\noindent Thus, synaptic plasticity is defined by: 
\begin{equation} \label{hebbianGrowthEqn}
\Delta w_{ab}(t) = \gamma f_a(t) f_b(t), \text{ where } \gamma \text{ is a growth parameter. } \\
\end{equation} 

Updates occur every timestep within a short window after odor puff onset (odor puffs are 0.2 sec long).
There are two layers of plastic synaptic weights, pre- and post-MB:  
AL$\rightarrow$MB ($M^{P,K},M^{Q,K}$), and MB$\rightarrow$ENs ($M^{K,E}$) .
Learning rate parameters of \textit{MothNet} were calibrated to match experimental effects of octopamine 
on PN firing rates and known moth learning speed (\textit{e.g.}4 - 8 trials to induce behavior modification) \cite{riffell2013}.
Synaptic weights had maximum allowed values, reached in about 5 to 15 training odor puffs depending on the \textit{MothNet} template used.
The version of  \textit{MothNet} we used did not decay unused synaptic weights. 
Training does not alter octopamine delivery strength matrices ($M^{O,\text{*}}$); that is, the neuromodulator channels are not plastic (unlike, for example, the case in \cite{grant}).  \\

\noindent\textit{Odor and octopamine injections:}~\\
Odors and octopamine are modeled as hamming windows.
The smooth leading and trailing edges ensures low stiffness of the dynamic ODEs, and allows 
a 10 mSec timestep to give accurate evolution of the SDEs in simulations.\\ 

\noindent \textit{Training:}~\\
Training on an odor consists of simultaneously applying puffs of the odor, injecting octopamine, and ``switching on" Hebbian growth.
Training with 5 to 10 odor puffs typically produces behavior change in live moths. 

\subsection{Governing equations and numerical schemes}
\label{sectionGoverningEqns}
Equations governing the time-stepped simulations of the MON system \cite{delahuntMoth1} are given in Eqns.\ref{eqnGovFirst} to \ref{eqnGovLast} and Table \ref{eqnDefnTable}.
The sigmoid function was piecewise-linear to speed simulations, though a standard sigmoid works as well. 

\begin{eqnarray}
\label{eqnGovFirst}
\tau_{\text{{\tiny R}}} \cdot d \bold{u}^{\text{{\tiny R}}}  &=&   f_{\text{{\tiny R}}} \big( \bold{u}^{\text{{\tiny R}}} , \bold{u}^{\text{{\tiny L}}} , \bold{u}^{\text{{\tiny S}}} ,  M^{\text{{\tiny L,R}}}, M^{\text{{\tiny S,R}}} , M^{\text{{\tiny O,R}}},o(t) \big)dt ~+~ d \bold{W}^{\text{{\tiny R}}} \\
\vspace{.3em}
\tau_{\text{{\tiny P}}} \cdot d \bold{u}^{\text{{\tiny P}}}  &=&   f_{\text{{\tiny P}}} \big( \bold{u}^{\text{{\tiny R}}} , \bold{u}^{\text{{\tiny P}}} , \bold{u}^{\text{{\tiny L}}}, M^{\text{{\tiny L,P}}}, M^{\text{{\tiny R,P}}} ,  M^{\text{{\tiny O,P}}},o(t) \big)dt  ~+~ d \bold{W}^{\text{{\tiny P}}} \\
\vspace{.3em}
\tau_{\text{{\tiny Q}}} \cdot d \bold{u}^{\text{{\tiny Q}}} &=&  f_{\text{{\tiny Q}}} \big( \bold{u}^{\text{{\tiny R}}} , \bold{u}^{\text{{\tiny Q}}} , \bold{u}^{\text{{\tiny L}}} , M^{\text{{\tiny L,Q}}}, M^{\text{{\tiny R,Q}}}, M^{\text{{\tiny O,Q}}},o(t) \big)dt  ~+~ d \bold{W}^{\text{{\tiny Q}}} \\
\vspace{.3em}
\tau_{\text{{\tiny L}}} \cdot d \bold{u}^{\text{{\tiny L}}} &=&  f_{\text{{\tiny L}}}\big ( \bold{u}^{\text{{\tiny R}}} , \bold{u}^{\text{{\tiny L}}},   M^{\text{{\tiny L,L}}}, M^{\text{{\tiny R,L}}}, M^{\text{{\tiny O,L}}},o(t) \big)dt  ~+~ d \bold{W}^{\text{{\tiny L}}} \\
\vspace{.3em}
\tau_{\text{{\tiny K}}} \cdot d \bold{u}^{\text{{\tiny K}}}  &=&  f_{\text{{\tiny K}}}\big ( \bold{u}^{\text{{\tiny P}}} , \bold{u}^{\text{{\tiny Q}}},  M^{\text{{\tiny P,K}}}, M^{\text{{\tiny Q,K}}} \big)dt ~+~ d \bold{W}^{\text{{\tiny K}}} \\  
\vspace{.3em}
\label{eqnGovLast}
\tau_{\text{{\tiny E}}} \cdot d \bold{u}^{\text{{\tiny E}}} &=&  f_{\text{{\tiny E}}}\big ( \bold{u}^{\text{{\tiny K}}} , \bold{u}^{\text{{\tiny E}}},  M^{\text{{\tiny K,E}}} \big)dt 
\end{eqnarray}
where
{\scriptsize
\begin{equation*}
\begin{cases}
f_{\text{{\tiny R}}}  =  - \bold{u}^{\text{{\tiny R}}}  + \text{sigmoid} \big [ -\big (I-  \alpha \cdot o(t) \cdot M^{\text{{\tiny O,R}}} \big ) 
M^{\text{{\tiny L,R}}} \hspace{.2em} {\bf{u}}^{\text{{\tiny L}}} ~+~ \big ( I + o(t)\cdot M^{\text{{\tiny O,R}}} \big) M^{\text{{\tiny S,R}}} \hspace{.2em}  \bf{u}^{\text{{\tiny S}}}
\big ]  \\
\vspace{.3em}
f_{\text{{\tiny P}}}  =  - \bold{u}^{\text{{\tiny P}}}  + \text{sigmoid} \big [ -\big (I- \alpha \cdot o(t) \cdot M^{\text{{\tiny O,P}}} \big )  M^{\text{{\tiny L,P}}}  \hspace{.2em}  {\bf{u}}^{\text{{\tiny L}}} 
~+~ \big ( I + o(t)\cdot M^{\text{{\tiny O,P}}} \big) M^{\text{{\tiny R,P}}} \hspace{.2em}  \bf{u}^{\text{{\tiny R}}} \big ]  \\
\vspace{.3em}
f_{\text{{\tiny Q}}}  =  - \bold{u}^{\text{{\tiny Q}}}  + \text{sigmoid} \big [ -\big (I- \alpha \cdot o(t) \cdot M^{\text{{\tiny O,Q}}} \big )  M^{\text{{\tiny L,Q}}}  \hspace{.2em}  {\bf{u}}^{\text{{\tiny L}}} 
~+~ \big ( I + o(t)\cdot M^{\text{{\tiny O,Q}}} \big) M^{\text{{\tiny R,Q}}} \hspace{.2em}  \bf{u}^{\text{{\tiny R}}} \big ] \\
\vspace{.3em}
f_{\text{{\tiny L}}}  =  - \bold{u}^{\text{{\tiny L}}}  + \text{sigmoid} \big [ -\big (I- \alpha \cdot o(t) \cdot M^{\text{{\tiny O,L}}} \big ) 
M^{\text{{\tiny L,L}}} \hspace{.2em}  {\bf{u}}^{\text{{\tiny L}}} ~+~ \big ( I + o(t)\cdot M^{\text{{\tiny O,L}}} \big) M^{\text{{\tiny R,L}}} \hspace{.2em}  \bf{u}^{\text{{\tiny R}}}
\big ]  \\
\vspace{.3em}
f_{\text{{\tiny K}}}  =  - \bold{u}^{\text{{\tiny K}}}  + \text{sigmoid} \big [ -\big ( {\bf{u}}^{\text{{\tiny D}}} + M^{\text{{\tiny Q,K}}} \hspace{.2em} {\bf{u}}^{\text{{\tiny Q}}} \big)  ~+~   M^{\text{{\tiny P,K}}} \hspace{.2em} \bf{u}^{\text{{\tiny P}}}
\big ]  \\
\vspace{.3em}
f_{\text{{\tiny E}}}  =  - \bold{u}^{\text{{\tiny E}}}  + M^{\text{{\tiny K,E}}} \hspace{.2em}  \bf{u}^{\text{{\tiny K}}}   
\end{cases}
\end{equation*}
}
\begin{table*}
 \label{eqnDefnTable}
\centering
\caption{List of model parameters and variables.} 
\vspace{1em}
\footnotesize
\begin{tabular}{ c c c l }
Symbol & Type & Size/Value & Description and Remarks
 \\ \hline
 \\
R & superscript & & Refers to the \textit{receptor neurons} subpopulation. \\
P & superscript & & Refers to the \textit{excitatory projection neurons} subpopulation. \\
Q & superscript & & Refers to the \textit{inhibitory projection neurons} subpopulation.  \\
L & superscript & & Refers to the \textit{lateral neurons} subpopulation. \\
K & superscript & & Refers to the \textit{kenyon cells} subpopulation.  \\
E & superscript & & Refers to the readout \textit{extrinsic neurons} subpopulation. \\
O & superscript & & Refers to the \textit{octopamine} neurotransmitter. \\
\\
$nG$ & scalar & 60 & Number of glomeruli in the antennal lobe. $^{*}$ \\
$nP$ & scalar & nG & Number of excitatory projection  neurons. \\ 
$nQ$ & scalar & variable & Number of inhibitory projection neurons. \\
$nK$ & scalar & 2000 & Number of kenyon cells. \\
$nE$ & scalar & 1 & Number of extrinsic neurons. \\
 
\\ 
 $\bold{u}^{\text{{\tiny R}}}$ & vector &  $nG \times 1$ & FRs of the  receptor neurons (RNs).\\
 $\bold{u}^{\text{{\tiny P}}}$ & vector &  $nP \times 1$ & FRs of the  excitatory projection neurons (PNs).\\
 $\bold{u}^{\text{{\tiny Q}}}$ & vector &  $nQ \times 1$ & FRs of the inhibitory projection neurons (QNs).\\
 $\bold{u}^{\text{{\tiny L}}}$ & vector &  $nG \times 1$ & FRs of the  inhibitory intra-AL neurons (LNs). \\
 $\bold{u}^{\text{{\tiny K}}}$ & vector &  $nK \times 1$ & FRs of the kenyon cells (KCs). Sparse. \\
 $\bold{u}^{\text{{\tiny E}}}$ & vector &  $nE \times 1$ & FRs of the  extrinsic neurons (ENs). \\
 \\ 
    
$M^{\text{{\tiny S,R}}}$ & matrix & $nG \times nS$ & Stimulus $\rightarrow \bold{u}^{\text{{\tiny R}}}$ connections.  \\
\\
$M^{\text{{\tiny O,R}}}$ & matrix & $nG \times nG$ & Octopamine $\rightarrow \bold{u}^{\text{{\tiny R}}}$ connections. Diagonal matrix.\\
$M^{\text{{\tiny O,L}}}$ & matrix & $nG \times nG$ & Octopamine $\rightarrow \bold{u}^{\text{{\tiny L}}}$ connections. Diagonal matrix. \\
\\
$M^{\text{{\tiny R,L}}}$ & matrix & $nG \times nG$ & Connection weights ~ $\bold{u}^{\text{{\tiny R}}} \rightarrow \bold{u}^{\text{{\tiny L}}}$.\\
$M^{\text{{\tiny R,P}}}$ & matrix & $nG \times nG$ & Connection weights ~ $\bold{u}^{\text{{\tiny R}}} \rightarrow \bold{u}^{\text{{\tiny P}}}$. Diagonal matrix.\\
$M^{\text{{\tiny R,Q}}}$ & matrix & $nQ \times nG$ & Connection weights ~ $\bold{u}^{\text{{\tiny R}}} \rightarrow \bold{u}^{\text{{\tiny Q}}}$.\\
$M^{\text{{\tiny P,K}}}$ & matrix & $nK \times nG$ & Connection weights ~ $\bold{u}^{\text{{\tiny P}}} \rightarrow \bold{u}^{\text{{\tiny K}}}$.\\
$M^{\text{{\tiny Q,K}}}$ & matrix & $nK \times nQ$ & Connection weights ~ $\bold{u}^{\text{{\tiny Q}}} \rightarrow \bold{u}^{\text{{\tiny K}}}$.\\
$M^{\text{{\tiny L,R}}}$ & matrix & $nG \times nG$ & Connection weights ~ $\bold{u}^{\text{{\tiny L}}} \rightarrow \bold{u}^{\text{{\tiny R}}}$. \\
$M^{\text{{\tiny L,P}}}$ & matrix & $nG \times nG$ & Connection weights ~ $\bold{u}^{\text{{\tiny L}}} \rightarrow \bold{u}^{\text{{\tiny P}}}$.\\
$M^{\text{{\tiny L,Q}}}$ & matrix & $nQ \times nG$& Connection weights ~ $\bold{u}^{\text{{\tiny L}}} \rightarrow \bold{u}^{\text{{\tiny Q}}}$. \\
$M^{\text{{\tiny L,L}}}$ & matrix & $nG \times nG$ & Connection weights ~ $\bold{u}^{\text{{\tiny L}}} \rightarrow \bold{u}^{\text{{\tiny L}}}$. \\
$M^{\text{{\tiny K,E}}}$ & matrix & $nE \times nK$ & Connection weights ~ $\bold{u}^{\text{{\tiny K}}} \rightarrow \bold{u}^{\text{{\tiny E}}}$. \\
\\
$o(t)$ & function &  0 or 1 & Flags when octopamine is active (typically during training).\\
$\gamma $ & scalar & 0.5 & Scaling factor for octopamine's effects on inhibition. $^{*}$ \\
\\
$\tau_{\text{{\tiny R}}}$ & scalar & &Time constants were set to ensure $\approx 1~sec$ response recovery time. \\
$\tau_{\text{{\tiny P}}}$ & scalar & & \\
$\tau_{\text{{\tiny Q}}}$ & scalar & & \\
$\tau_{\text{{\tiny L}}}$ & scalar & & \\
$\tau_{\text{{\tiny K}}}$ & scalar & & \\
$\tau_{\text{{\tiny E}}}$ & scalar & & \\\\
\hline
\end{tabular}
\\
{\scriptsize $^{*}$  Each glomerulus receives one (collective) RN unit and one octopamine input, and initiates five PNs, 1 to 5 QNs, and one LN. \\
$^{*}$ Octopamine decreases the response to inhibition less than it increases the response to excitation}\\
\label{NetSymbols}
\end{table*}

\noindent \textit{Discretization:} ~\\
We used the standard Euler-Maruyama (EM) forward-step method for SDEs  \cite{higham2001}.
\begin{enumerate}
\item  Euler (noise-free): $x_{n + 1} = x_{n} + \Delta t f(x_{n} )$ 
\item  Euler-Maruyama:  $x_{n + 1} = x_{n} + \Delta t f(x_{n} ) + \epsilon \text{ randn(0,1)} \sqrt{\Delta t}$, where $\epsilon$ controls the noise intensity.  
\end{enumerate}

\noindent\textit{Convergence:} ~\\
The timestep $\Delta t$ was chosen such that noise-free EM evolution gives the same timecourses as Runge-Kutta (4th order), via Matlab's ode45 function.
$\Delta t = 10$ mSec suffices to match EM evolution to RK in noise-free moths.
Values of $\Delta t \leq 20$ mSec gives equivalent simulations in moths with AL noise calibrated to match \textit{in vivo} data. 
Values of $\Delta t \geq 40$ mSec show differences in evolution outcomes given AL noise.


\section{ Appendix: \textit{P}-values} 

We give \textit{P}-values for Figures \ref{varyQnResultsPanel}, \ref{varyALNoiseResultsPanel}, and  \ref{alNoiseMuVsMuPlusSigma}, as evidence of meaningful differences between injury mitigation associated with amount of AL noise, or number of parallel inhibitory connections (QN:PN ratio). 
We avoid the term ``significant" in association with \textit{P}-values, following the arguments in \cite{rePvalues}.
\textit{P}-values here are calculated as the likelihood of the mean of one distribution occurring given another (implicitly null) distribution. 
\textit{P}-values for post-injury responses had much higher \textit{N}, because moth instances with any number of training puffs could be combined (injury occurs before training in the simulations, so amount of training is irrelevant).

\subsection{Effects of AL noise}

These tables refer to Figs.\ref{varyALNoiseResultsPanel} (A, B) and \ref{alNoiseMuVsMuPlusSigma} (A - D).
\textit{P}-values are given in four tables for \textit{top-scoring} responses and for \textit{average} responses, both post-\textit{injury} and post-\textit{training}.
Two tables then give \textit{P}-values for top-scoring \textit{vs} average responses,  post-injury and post-training.
They are calculated from the same set of simulations as the figures.

The key points are that \textit{(i)}  the increased protective effect due to increased AL noise was meaningful on the top-scoring responses; \textit{(ii)} this protective effect was noticeably lower for average responses; and \textit{(iii)} this protective effect was meaningfully greater for top-scoring than for average responses, both post-injury and post-training.

\begin{table}[h!]
\caption{ \normalfont{\textit{P}-values for  normalized post-\textit{injury} effects of increased AL noise, on \textit{top-scoring} responses (see Fig.\ref{varyALNoiseResultsPanel} A).  
\textit{P}-values are for a given AL noise vs AL noise = 0.67 (\textit{i.e.} the middle value).
``0+'' indicates \textit{P}-value $<$ 0.0005.
$N$ = 98 to 112 (mean = 107).
}}
\begin{center}
{\renewcommand{\arraystretch}{1.2}
\begin{tabular}{|c|c|c|c|c|c|}
\hline  
Injury level  & \multicolumn{5}{|c|}{AL noise level}     \\
(\%) &0 &   0.33  & 0.67 & 1.0  & 1.33  \\ \hline 
 10 & 0+    &     0.015      &      0.5      &     0.15      &    0.006     \\
  20& 0+      &        0+      &      0.5      &    0.161    &      0.009     \\
 30 & 0+      &    0.001     &      0.5      &    0.027      &        0+     \\
 40 & 0+      &    0.363      &     0.5     &     0.001       &       0+     \\
 50 & 0+      &        0     &       0.5      &    0.036     &     0.013     \\
 60 & 0+      &    0.002    &       0.5      &        0+       &       0+     \\
\hline
\end{tabular}
}
\label{tableALNoiseTopPostInjury}
\end{center}
\end{table}

\begin{table}[h!]
\caption{ \normalfont{\textit{P}-values for  normalized post-\textit{injury} effects of increased AL noise on \textit{average} responses (see Fig.\ref{alNoiseMuVsMuPlusSigma} C).  
\textit{P}-values are much larger than for those corresponding to top-scoring responses (cf Table \ref{tableALNoiseTopPostInjury}), indicating weaker protective effects.
\textit{P}-values are for a given AL noise vs AL noise = 0.67 (\textit{i.e.} the middle value).
``0+'' indicates \textit{P}-value $<$ 0.0005.
$N$ = 98 to 112 (mean = 107).
} }
\begin{center}
{\renewcommand{\arraystretch}{1.2}
\begin{tabular}{|c|c|c|c|c|c|}
\hline  
Injury level  & \multicolumn{5}{|c|}{AL noise level}     \\
 (\%)&0 &   0.33  & 0.67 & 1.0  & 1.33  \\ \hline
10 &   0.032    &     0.057     &      0.5     &     0.355      &    0.267    \\
20  & 0.026    &     0.001       &     0.5     &     0.495      &    0.061    \\
 30 & 0.028     &     0.022     &      0.5     &     0.225      &    0.008    \\
 40 & 0.056     &     0.703      &      0.5     &     0.023      &        0    \\
 50 & 0.033      &    0.018      &      0.5     &     0.132      &    0.087    \\
 60 & 0.003     &     0.045     &      0.5        &      0+          &    0+     \\
\hline
\end{tabular}
}
\label{tableALNoiseAvePostInjury}
\end{center}
\end{table}

\begin{table}[h!]
\caption{ \normalfont{\textit{P}-values for  normalized post-\textit{training} effects of increased AL noise on \textit{top-scoring} responses (see Fig.\ref{varyALNoiseResultsPanel} B).  
\textit{P}-values are for a given AL noise vs AL noise = 0.67 (\textit{i.e.} the middle value).
``0+'' indicates \textit{P}-value $<$ 0.0005.
$N$ = 31 to 40 (mean = 35). 
}}
\begin{center}
{\renewcommand{\arraystretch}{1.2}
\begin{tabular}{|c|c|c|c|c|c|}
\hline  
Injury level  & \multicolumn{5}{|c|}{AL noise level}     \\
 (\%)&0 &   0.33  & 0.67 & 1.0  & 1.33  \\ \hline
10 &      0+     &     0.003      &      0.5     &     0.545      &    0.789     \\
 20  &    0+      &    0.001       &     0.5     &     0.866      &    0.525     \\
30  & 0.001     &     0.004     &       0.5     &      0.29       &   0.183     \\
40 & 0.008      &    0.116     &      0.5      &    0.404     &     0.006     \\
 50  &  0.1      &    0.086      &     0.5      &    0.072      &    0.007     \\
60  & 0.034    &     0.011      &     0.5     &     0.011     &     0.001     \\
\hline
\end{tabular}
}
\label{tableALNoiseTopPostTrain}
\end{center}
\end{table}

\begin{table}[h!]
\caption{ \normalfont{\textit{P}-values for normalized post-\textit{training} effects of increased AL noise on \textit{average} responses (see Fig.\ref{alNoiseMuVsMuPlusSigma} D).  
\textit{P}-values are larger than for those corresponding to top-scoring responses (cf Table \ref{tableALNoiseTopPostTrain}), indicating weaker protective effects.
\textit{P}-values are for a given AL noise vs AL noise = 0.67 (\textit{i.e.} the middle value).
$N$ = 31 to 40 (mean = 35). 
} }
\begin{center}
{\renewcommand{\arraystretch}{1.2}
\begin{tabular}{|c|c|c|c|c|c|}
\hline  
Injury level  & \multicolumn{5}{|c|}{AL noise level}     \\
 (\%)&0 &   0.33  & 0.67 & 1.0  & 1.33  \\ \hline
10   & 0.001   &   0.035     &       0.5      &    0.833     &     0.913     \\
 20 & 0.031     &     0.032     &      0.5     &      0.89      &    0.715     \\
  30& 0.049     &     0.025    &       0.5     &     0.755     &     0.534     \\
  40 & 0.33     &    0.271      &      0.5     &      0.67      &    0.066     \\
 50 & 0.548     &      0.16       &     0.5      &    0.278      &    0.078     \\
  60& 0.344     &     0.058    &       0.5      &    0.021     &     0.005     \\
\hline
\end{tabular}
}
\label{tableALNoiseAvePostTrain}
\end{center}
\end{table}

\clearpage

\begin{table}[h!]
\caption{ \normalfont{ \textit{P}-values for comparison of AL noise's effects on  \textit{top-scoring}  responses \textit{vs} \textit{average} responses, post-\textit{injury} (see Fig.\ref{alNoiseMuVsMuPlusSigma} A\,\textit{vs}\,C).}  
``0+'' indicates \textit{P}-value $<$ 0.0005.
$N$ = 98 to 112 (mean = 106).
}
\begin{center}
{\renewcommand{\arraystretch}{1.2}
\begin{tabular}{|c|c|c|c|c|c|}
\hline  
Injury level  & \multicolumn{5}{|c|}{AL noise level}     \\
 (\%)&0 &   0.33  & 0.67 & 1.0  & 1.33  \\ \hline
 10&  0.995      &    0.667    &      0.224      &    0.064     &     0.157  \\
  20& 0.903     &     0.447      &    0.078     &     0.182     &      0.06  \\
  30& 0.787      &    0.395      &    0.211     &     0.015    &      0.022  \\
  40& 0.747     &     0.243     &      0.12     &     0.004     &     0.009  \\
  50& 0.676     &     0.242      &    0.154      &    0.024     &     0.002  \\
  60& 0.613      &    0.256      &    0.079      &     0.01      &        0+   \\
\hline
\end{tabular}
}
\label{tableALNoiseTopVsAllPostInjury}
\end{center}
\end{table}

\begin{table}[h!]
\caption{ \normalfont{ \textit{P}-values for comparison of AL noise's effects on  \textit{top-scoring} responses \textit{vs} \textit{average} responses, post-\textit{training} (see Fig.\ref{alNoiseMuVsMuPlusSigma} B\,\textit{vs}\,D).}  
``0+'' indicates \textit{P}-value $<$ 0.0005.
$N$ = 31 to 40 (mean = 35). 
  }
\begin{center}
{\renewcommand{\arraystretch}{1.2}
\begin{tabular}{|c|c|c|c|c|c|}
\hline  
Injury level & \multicolumn{5}{|c|}{AL noise level}     \\
 (\%)&0 &   0.33  & 0.67 & 1.0  & 1.33  \\ \hline
10 &   1     &   0.777    &    0.102     &   0.004     &   0.044     \\
20&0.988     &   0.406     &   0.009    &    0.059     &   0.004     \\
 30&0.92     &   0.323     &   0.084      &      0+       &     0+     \\
40&0.874    &    0.111      &   0.02       &     0+       &     0+     \\
50&0.799     &   0.114     &   0.038       &     0+      &      0+     \\
60&0.699     &   0.123    &   0.008       &     0+       &     0+     \\
\hline
\end{tabular}
}
\label{tableALNoiseTopVsAllPostTrain}
\end{center}
\end{table}


\subsection{Effects of QN ratio}

Tables \ref{tableQNsPostInjury} and \ref{tableQNsPostTrain} give \textit{P}-values for each \{injury level, \#QNs\}, compared to the next lower (adjacent) \#QNs (same injury level). 
These tables refer to Fig.\ref{varyQnResultsPanel} (A, B).
They are calculated from a different set of simulations, which however had identical parameters and gave the same overall results.
In general the differences are meaningful, indicating that increasing the number of parallel inhibitory neurons improves injury mitigation properties.

\begin{table}[h!]
\caption{ \normalfont{ \textit{P}-values for  normalized post-injury responses (see Fig.\ref{varyQnResultsPanel} A).}  
\textit{P}-values are for \textit{adjacent} QN numbers. 
``0+'' indicates \textit{P}-value $<$ 0.0005.
\textit{n} = 66 to 97 (mean = 79).
}
\begin{center}
{\renewcommand{\arraystretch}{1.2}
\begin{tabular}{|c|c|c|c|c|}
\hline  
Injury level  & \multicolumn{4}{|c|}{number of QNs} \\
 (\%) &   2  & 4  & 5  & 7   \\ \hline
0 &    0.556   &     0.306   &    0.813   &   0.955\\
10   &     0.007 &     0.01  &    0.286    &    0.06\\
20    &       0+    &    0.003    &    0.027     &       0+\\
30     &      0+     &       0+    &    0.105     &       0+\\
40      &      0+      &      0+    &    0.025     &       0+\\
50    &    0.015    &        0+     &   0.091    &        0+\\
60    &        0+       &     0+    &       0+    &        0+\\
\hline
\end{tabular}
}
\label{tableQNsPostInjury}
\end{center}
\end{table}

\begin{table}[h!]
\caption{ \normalfont{\textit{P}-values for normalized post-training responses (see Fig.\ref{varyQnResultsPanel} B).}  
\textit{P}-values are for \textit{adjacent} (next-lower) QN numbers. 
``0+'' indicates \textit{P}-value $<$ 0.0005.
\textit{n} = 31 to 62 (mean = 40).}
\begin{center}
{\renewcommand{\arraystretch}{1.2}
\begin{tabular}{|c|c|c|c|c|}
\hline  
Injury level   & \multicolumn{4}{|c|}{number of QNs} \\
  (\%)&   2  & 4  & 5  & 7   \\ \hline
0    &   0.316   &    0.365    &   0.867     &  0.162 \\
 10     &      0+   &     0.02    &   0.141   &    0.093 \\
20     &     0+    &   0.001   &    0.035    &       0+ \\
  30      &    0+      &     0+    &   0.349     &      0+ \\
  40     &      0+      &     0+   &    0.055    &       0+ \\
 50    &   0.532      &     0+   &    0.004    &       0+ \\
 60      &     0+      &     0+      &     0+     &      0+ \\
\hline
\end{tabular}
}
\label{tableQNsPostTrain}
\end{center}
\end{table}


\section*{Acknowledgements} 
CBD was partially supported by the Swartz Foundation.\\
JNK acknowledges support from the Air Force Office of Scientific Research (FA9550-19-1-0011).

%

 \clearpage

%
%
%
%
\bibliographystyle{spmpsci}  
 
\bibliography{mothBibliography_june2019}

\begin{thebibliography}{10}
\providecommand{\url}[1]{{#1}}
\providecommand{\urlprefix}{URL }
\expandafter\ifx\csname urlstyle\endcsname\relax
  \providecommand{\doi}[1]{DOI~\discretionary{}{}{}#1}\else
  \providecommand{\doi}{DOI~\discretionary{}{}{}\begingroup
  \urlstyle{rm}\Url}\fi

\bibitem{bazhenovStopfer2010}
Bazhenov, M., Stopfer, M.: Forward and back: Motifs of inhibition in olfactory
  processing.
\newblock Neuron \textbf{67}(3), 357 -- 358 (2010).
\newblock \doi{http://dx.doi.org/10.1016/j.neuron.2010.07.023}.
\newblock
  \urlprefix\url{http://www.sciencedirect.com/science/article/pii/S0896627310005842}

\bibitem{bhandawat2007}
Bhandawat, V., Olsen, S.R., Gouwens, N.W., Schlief, M.L., Wilson, R.I.: Sensory
  processing in the {D}rosophila antennal lobe increases reliability and
  separability of ensemble odor representations.
\newblock Nature Neuroscience \textbf{10}, 1474--1482 (2007).
\newblock
  \urlprefix\url{http://www.nature.com/neuro/journal/v10/n11/full/nn1976.html}

\bibitem{campbell2013}
Campbell, R., Honegger, K., Qin, H., Li, W., Demir, E., Turner, G.: Imaging a
  population code for odor identity in the {D}rosophila mushroom body.
\newblock Journal of Neuroscience \textbf{33}(25), 10,568--81 (2013).
\newblock \doi{10.1523/JNEUROSCI.0682-12.2013}

\bibitem{campbellMushroomBody}
Campbell, R.A., Turner, G.C.: The mushroom body.
\newblock Current Biology \textbf{20}(1), R11 -- R12 (2010).
\newblock \doi{https://doi.org/10.1016/j.cub.2009.10.031}.
\newblock
  \urlprefix\url{http://www.sciencedirect.com/science/article/pii/S096098220901851X}

\bibitem{cassenaer}
Cassenaer, S., Laurent, G.: Hebbian stdp in mushroom bodies facilitates the
  synchronous flow of olfactory information in locusts.
\newblock Nature \textbf{448}, 709 EP -- (2007).
\newblock \urlprefix\url{http://dx.doi.org/10.1038/nature05973}

\bibitem{dayan2001}
Dayan, P., Abbott, L.F.: Theoretical Neuroscience: Computational and
  Mathematical Modeling of Neural Systems.
\newblock The MIT Press (2005)

\bibitem{delahuntMothInjuryCodebase}
Delahunt, C.B.: Codebase for moth neural injury simulations (2018).
\newblock \urlprefix\url{https://github.com/charlesDelahunt/BuiltToLast}

\bibitem{delahuntMoth1}
Delahunt, C.B., Riffell, J.A., Kutz, J.N.: Biological mechanisms for learning:
  A computational model of olfactory learning in the {M}anduca sexta moth, with
  applications to neural nets.
\newblock Frontiers in Computational Neuroscience \textbf{12}, 102 (2018).
\newblock \doi{10.3389/fncom.2018.00102}.
\newblock
  \urlprefix\url{https://www.frontiersin.org/article/10.3389/fncom.2018.00102}

\bibitem{Eisthen2002}
Eisthen, H.L.: Why are olfactory systems of different animals so similar?
\newblock Brain, Behavior and Evolution \textbf{59}, 273--293 (2002)

\bibitem{galizia2014}
Galizia, C.G.: Olfactory coding in the insect brain: data and conjectures.
\newblock European Journal of Neuroscience \textbf{39}(11), 1784--1795 (2014).
\newblock \doi{10.1111/ejn.12558}.
\newblock \urlprefix\url{http://dx.doi.org/10.1111/ejn.12558}

\bibitem{ganguli2012}
Ganguli, S., Sompolinsky, H.: Compressed sensing, sparsity, and dimensionality
  in neuronal information processing and data analysis.
\newblock Annual Review of Neuroscience \textbf{35}(1), 485--508 (2012).
\newblock \doi{10.1146/annurev-neuro-062111-150410}.
\newblock \urlprefix\url{https://doi.org/10.1146/annurev-neuro-062111-150410}.
\newblock PMID: 22483042

\bibitem{grant}
Grant, W.S., Tanner, J., Itti, L.: Biologically plausible learning in neural
  networks with modulatory feedback.
\newblock Neural Networks \textbf{88}(Supplement C), 32 -- 48 (2017).
\newblock \doi{https://doi.org/10.1016/j.neunet.2017.01.007}.
\newblock
  \urlprefix\url{http://www.sciencedirect.com/science/article/pii/S0893608017300072}

\bibitem{guptaStopfer2012}
Gupta, N., Stopfer, M.: Functional analysis of a higher olfactory center, the
  lateral horn.
\newblock Journal of Neuroscience \textbf{32}(24), 8138--8148 (2012).
\newblock \doi{10.1523/JNEUROSCI.1066-12.2012}.
\newblock \urlprefix\url{http://www.jneurosci.org/content/32/24/8138}

\bibitem{hammer1995}
Hammer, M., Menzel, R.: Learning and memory in the honeybee.
\newblock Journal of Neuroscience \textbf{15}(3), 1617--1630 (1995).
\newblock \urlprefix\url{http://www.jneurosci.org/content/15/3/1617}

\bibitem{hammerMenzel1998}
Hammer, M., Menzel, R.: Multiple sites of associative odor learning as revealed
  by local brain microinjections of octopamine in honeybees.
\newblock Learn Mem \textbf{5}(1), 146--156 (1998).
\newblock \urlprefix\url{http://www.ncbi.nlm.nih.gov/pmc/articles/PMC311245/}.
\newblock 10454379[pmid]

\bibitem{hebb}
Hebb, D.O.: The organization of behavior : a neuropsychological theory.
\newblock Wiley New York (1949)

\bibitem{hige2015}
Hige, T., Aso, Y., Rubin, G.M., Turner, G.C.: Plasticity-driven
  individualization of olfactory coding in mushroom body output neurons.
\newblock Nature \textbf{526}, 258 EP -- (2015).
\newblock \urlprefix\url{http://dx.doi.org/10.1038/nature15396}

\bibitem{higginson}
Higginson, A.D., Barnard, C.J., Tofilski, A., Medina, L., Ratnieks, F.:
  Experimental wing damage affects foraging effort and foraging distance in
  honeybees {A}pis mellifera.
\newblock Psyche  (2011).
\newblock \urlprefix\url{http://dx.doi.org/10.1155/2011/419793}

\bibitem{higham2001}
Higham., D.J.: An algorithmic introduction to numerical simulation of
  stochastic differential equations.
\newblock SIAM Rev. \textbf{43}(3), 525--546 (2001).
\newblock \doi{10.1137/S0036144500378302}.
\newblock \urlprefix\url{http://dx.doi.org/10.1137/S0036144500378302}

\bibitem{honeggerTurner2011}
Honegger, K.S., Campbell, R.A.A., Turner, G.C.: Cellular-resolution population
  imaging reveals robust sparse coding in the {D}rosophila mushroom body.
\newblock Journal of Neuroscience \textbf{31}(33), 11,772--11,785 (2011).
\newblock \doi{10.1523/JNEUROSCI.1099-11.2011}.
\newblock \urlprefix\url{http://www.jneurosci.org/content/31/33/11772}

\bibitem{Klambt2009}
Klambt, C.: Modes and regulation of glial migration in vertebrates and
  invertebrates.
\newblock Nature Reviews Neurosciencevolume \textbf{10}, 769--779 (2009)

\bibitem{Kunert2017}
Kunert, J.M., Maia, P.D., Kutz, J.N.: Functionality and robustness of injured
  connectomic dynamics in c. elegans: Linking behavioral deficits to neural
  circuit damage.
\newblock PLOS Computational Biology \textbf{13}(1), 1--21 (2017).
\newblock \doi{10.1371/journal.pcbi.1005261}.
\newblock \urlprefix\url{https://doi.org/10.1371/journal.pcbi.1005261}

\bibitem{kvello2009}
Kvello, P., Lofaldli, B., Rybak, J., Menzel, R., Mustaparta, H.: Digital,
  three-dimensional average shaped atlas of the heliothis virescens brain with
  integrated gustatory and olfactory neurons.
\newblock Frontiers in Systems Neuroscience \textbf{3}, 14 (2009).
\newblock \doi{10.3389/neuro.06.014.2009}.
\newblock
  \urlprefix\url{https://www.frontiersin.org/article/10.3389/neuro.06.014.2009}

\bibitem{lin2014}
Lin, A.C., Bygrave, A.M., de~Calignon, A., Lee, T., Miesenb{\"o}ck, G.: Sparse,
  decorrelated odor coding in the mushroom body enhances learned odor
  discrimination.
\newblock Nature Neuroscience \textbf{17}, 559 EP -- (2014).
\newblock \urlprefix\url{http://dx.doi.org/10.1038/nn.3660}

\bibitem{Lusch2018}
Lusch, B., Weholt, J., Maia, P.D., Kutz, J.N.: Modeling cognitive deficits
  following neurodegenerative diseases and traumatic brain injuries with deep
  convolutional neural networks.
\newblock Frontiers in Neuroscience \textbf{123}, 154--164 (2018)

\bibitem{lofaldi2010}
Løfaldli, B., Kvello, P., Mustaparta, H.: Integration of the antennal lobe
  glomeruli and three projection neurons in the standard brain atlas of the
  moth heliothis virescens.
\newblock Frontiers in Systems Neuroscience \textbf{4}, 5 (2010).
\newblock \doi{10.3389/neuro.06.005.2010}.
\newblock
  \urlprefix\url{https://www.frontiersin.org/article/10.3389/neuro.06.005.2010}

\bibitem{Maia2015}
Maia, P.D., Hemphill, M.A., Zehnder, B., Zhang, C., Parker, K.K., Kutz, J.N.:
  Diagnostic tools for evaluating the impact of focal axonal swellings arising
  in neurodegenerative diseases and/or traumatic brain injury.
\newblock Journal of Neuroscience Methods \textbf{253}, 233--243 (2015)

\bibitem{Maia2014_2}
Maia, P.D., Kutz, J.N.: Compromised axonal functionality after
  neurodegeneration, concussion and/or traumatic brain injury.
\newblock Journal of Computational Neuroscience \textbf{27}, 317--332 (2014)

\bibitem{maiaCompromisedAxonal}
Maia, P.D., Kutz, J.N.: Compromised axonal functionality after
  neurodegeneration, concussion and/or traumatic brain injury.
\newblock Journal of Computational Neuroscience \textbf{37}(2), 317--332
  (2014).
\newblock \doi{10.1007/s10827-014-0504-x}.
\newblock \urlprefix\url{https://doi.org/10.1007/s10827-014-0504-x}

\bibitem{Maia2014_1}
Maia, P.D., Kutz, J.N.: Identifying critical regions for spike propagation in
  axon segments.
\newblock Journal of Computational Neuroscience \textbf{36}(2), 141--155 (2014)

\bibitem{maiaReactionTime}
Maia, P.D., Kutz, J.N.: Reaction time impairments in decision-making networks
  as a diagnostic marker for traumatic brain injuries and neurological
  diseases.
\newblock Journal of Computational Neuroscience \textbf{42}, 323--347 (2017)

\bibitem{Maia2019}
Maia, P.D., Raj, A., Kutz, J.N.: Slow-gamma frequencies are optimally guarded
  against effects of neurodegenerative diseases and traumatic brain injuries.
\newblock Journal of Computational Neuroscience \textbf{47}, 1--16 (2019)

\bibitem{martin2011}
Martin, J.P., Beyerlein, A., Dacks, A.M., Reisenman, C.E., Riffell, J.A., Lei,
  H., Hildebrand, J.G.: The neurobiology of insect olfaction: Sensory
  processing in a comparative context.
\newblock Progress in Neurobiology \textbf{95}(3), 427 -- 447 (2011).
\newblock \doi{https://doi.org/10.1016/j.pneurobio.2011.09.007}.
\newblock
  \urlprefix\url{http://www.sciencedirect.com/science/article/pii/S0301008211001742}

\bibitem{masse2009}
Masse, N.Y., Turner, G.C., Jefferis, G.S.: Olfactory information processing in
  {D}rosophila.
\newblock Current Biology \textbf{19}(16), R700 -- R713 (2009).
\newblock \doi{https://doi.org/10.1016/j.cub.2009.06.026}.
\newblock
  \urlprefix\url{http://www.sciencedirect.com/science/article/pii/S0960982209013013}

\bibitem{papadopoulou2011}
Papadopoulou, M., Cassenaer, S., Nowotny, T., Laurent, G.: Normalization for
  sparse encoding of odors by a wide-field interneuron.
\newblock Science \textbf{332}(6030), 721--725 (2011).
\newblock \doi{10.1126/science.1201835}.
\newblock \urlprefix\url{https://science.sciencemag.org/content/332/6030/721}

\bibitem{perisse2013}
Perisse, E., Burke, C., Huetteroth, W., Waddell, S.: Shocking revelations and
  saccharin sweetness in the study of {D}rosophila olfactory memory.
\newblock Curr Biol \textbf{23}(17), R752--R763 (2013).
\newblock \doi{10.1016/j.cub.2013.07.060}.
\newblock \urlprefix\url{http://www.ncbi.nlm.nih.gov/pmc/articles/PMC3770896/}.
\newblock S0960-9822(13)00921-4[PII], 24028959[pmid]

\bibitem{pouget2014}
Pouget, A., Narain, C.: A conversation with alexandre pouget.
\newblock Cold Spring Harb Symp Quant Biol 2014 \textbf{79}, 285--287 (2014).
\newblock \doi{doi:10.1101/sqb.2014.79.14}

\bibitem{riffell2013}
Riffell, J.A., Lei, H., Abrell, L., Hildebrand, J.G.: Neural basis of a
  pollinator{\textquoteright}s buffet: Olfactory specialization and learning in
  {M}anduca sexta.
\newblock Science  (2012).
\newblock \doi{10.1126/science.1225483}.
\newblock
  \urlprefix\url{http://science.sciencemag.org/content/early/2012/12/05/science.1225483}

\bibitem{roberts2015}
Roberts, J.C., Cartar, R.V.: Shape of wing wear fails to affect load lifting in
  common eastern bumble bees ({B}ombus impatiens) with experimental wing wear.
\newblock Canadian Journal of Zoology \textbf{93}(7), 531--537 (2015).
\newblock \doi{10.1139/cjz-2014-0317}.
\newblock \urlprefix\url{https://doi.org/10.1139/cjz-2014-0317}

\bibitem{roelfsema2018}
Roelfsema, P.R., Holtmaat, A.: Control of synaptic plasticity in deep cortical
  networks.
\newblock Nature Reviews Neuroscience \textbf{19}, 166 EP -- (2018).
\newblock \urlprefix\url{http://dx.doi.org/10.1038/nrn.2018.6}.
\newblock Review Article

\bibitem{Rudy2016}
Rudy, S., Maia, P.D., Kutz, J.N.: Cognitive and behavioral deficits arising
  from neurodegeneration and traumatic brain injury: a model for the underlying
  role of focal axonal swellings in neuronal networks with plasticity.
\newblock Journal of Systems and Integrative Neuroscience  (2016)

\bibitem{Smith1994}
Smith, K.J.: Conduction properties of central demyelinated and remyelinated
  axons, and their relation to symptom production in demyelinating disorders.
\newblock Eye \textbf{8}, 224--237 (1994)

\bibitem{wang}
Wang, J., Hamm, R.J., Povlishock, J.T.: Traumatic axonal injury in the optic
  nerve: Evidence for axonal swelling, disconnection, dieback, and
  reorganization.
\newblock J Neurotrauma \textbf{28}(7), 1185--1198 (2011).
\newblock \doi{10.1089/neu.2011.1756[PII]}.
\newblock \urlprefix\url{http://www.ncbi.nlm.nih.gov/pmc/articles/PMC3136743/}.
\newblock 21506725[pmid]

\bibitem{rePvalues}
Wasserstein, R.L., Lazar, N.A.: The {ASA} statement on p-values: Context,
  process, and purpose.
\newblock The American Statistician \textbf{70}(2), 129--133 (2016).
\newblock \doi{10.1080/00031305.2016.1154108}.
\newblock \urlprefix\url{https://doi.org/10.1080/00031305.2016.1154108}

\bibitem{Weber2017}
Weber, M., Maia, P.D., Kutz, J.N.: Estimating memory deterioration rates
  following neurodegeneration and traumatic brain injuries in a hopfield
  network model.
\newblock Frontiers in Neuroscience \textbf{11:623} (2017)

\bibitem{wilson2008}
Wilson, R.I.: Neural and behavioral mechanisms of olfactory perception.
\newblock Current Opinion in Neurobiology \textbf{18}(4), 408 -- 412 (2008).
\newblock \doi{https://doi.org/10.1016/j.conb.2008.08.015}.
\newblock
  \urlprefix\url{http://www.sciencedirect.com/science/article/pii/S0959438808000883}.
\newblock Sensory systems

\bibitem{wilson2005}
Wilson, R.I., Laurent, G.: Role of gabaergic inhibition in shaping odor-evoked
  spatiotemporal patterns in the {D}rosophila antennal lobe.
\newblock Journal of Neuroscience \textbf{25}(40), 9069--9079 (2005).
\newblock \doi{10.1523/JNEUROSCI.2070-05.2005}.
\newblock \urlprefix\url{http://www.jneurosci.org/content/25/40/9069}

\end{thebibliography}

\end{document}